\documentclass[aps,prx,amsmath,superscriptaddress,twocolumn,nofootinbib,longbibliography]{revtex4-2}

\usepackage{graphicx}
\usepackage{amsmath}
\usepackage{amsfonts}
\usepackage{amssymb}
\usepackage{braket}
\usepackage{bm}
\usepackage{multirow}
\usepackage{color}
\usepackage[normalem]{ulem}
\usepackage{mathrsfs}
\usepackage{mathtools}
\usepackage{float}
\usepackage[textsize=small]{todonotes}

\usepackage[italicdiff]{physics}

\makeatletter

\newcommand{\Rmnum}[1]{\expandafter\@slowromancap\romannumeral #1@}
\makeatother

\usepackage[breaklinks]{hyperref}
\hypersetup{colorlinks=true, linkcolor=blue, citecolor=blue, filecolor=blue, urlcolor=blue}

\AtBeginDocument{%
    \newwrite\bibnotes
    \def\bibnotesext{Notes.bib}
    \immediate\openout\bibnotes=\jobname\bibnotesext
    \immediate\write\bibnotes{@CONTROL{REVTEX41Control}}
    \immediate\write\bibnotes{@CONTROL{%
    apsrev41Control,author="08",editor="1",pages="1",title="0",year="1"}}
     \if@filesw
     \immediate\write\@auxout{\string\citation{apsrev41Control}}%
    \fi
}%

\begin{document}

\title{A $\mathbb{Z}_3$ quantum double in a superconducting wire array}

\author{Zhi-Cheng Yang}
\affiliation{Joint Quantum Institute, University of Maryland, College Park, Maryland 20742, USA}
\affiliation{Joint Center for Quantum Information and Computer Science, University of Maryland, College Park, Maryland 20742, USA}

\author{Dmitry Green}
\affiliation{AppliedTQC.com, ResearchPULSE LLC, New York, NY 10065, USA}

\author{Hongji Yu}
\affiliation{Physics Department, Boston University, Boston, MA 02215, USA}

\author{Claudio Chamon}
\affiliation{Physics Department, Boston University, Boston, MA 02215, USA}

\date{\today}

\begin{abstract}

We show that a $\mathbb{Z}_3$ quantum double can be realized in an
array of superconducting wires coupled via Josephson junctions. With a
suitably chosen magnetic flux threading the system, the inter-wire
Josephson couplings take the form of a complex Hadamard matrix, which
possesses combinatorial gauge symmetry --- a local $\mathbb{Z}_3$
symmetry involving permutations and shifts by $\pm 2\pi/3$ of the
superconducting phases. The sign of the star potential resulting from
the Josephson energy is inverted in this physical realization, leading
to a massive degeneracy in the non-zero flux sectors. A dimerization
pattern encoded in the capacitances of the array lifts up these
degeneracies, resulting in a $\mathbb{Z}_3$ topologically ordered
state. Moreover, this dimerization pattern leads to a larger effective
vison gap as compared to the canonical case with the usual
(uninverted) star term. We further show that our model maps to a
quantum three-state Potts model under a duality transformation. We
argue, using a combination of bosonization and mean field theory, that
altering the dimerization pattern of the capacitances leads to a
transition from the $\mathbb{Z}_3$ topological phase into a quantum
XY-ordered phase. Our work highlights that combinatorial gauge
symmetry can serve as a design principle to build quantum double
models using systems with realistic interactions.
\end{abstract}

\maketitle


\section{Introduction}

The identification of possible experimental realizations of
topologically ordered states of matter remains a central problem in
condensed matter physics. The fractional quantum Hall (FQH)
effects~\cite{Tsui-etal, Laughlin} are the quintessential and best
characterized topological states. Both the fractional
charge~\cite{Glattli, Reznikov} and, more recently, the fractional
statistics~\cite{Manfra, Bartolomei2020} of the quasiparticle
excitations of Abelian FQH states have been experimentally
measured. In addition to their fundamental importance, topological
phases such as those associated with non-Abelian FQH states have
potential application to topological quantum computation.

Underlying all qubits that are based on topological ordered
states~\cite{Wen1990a} are quantum liquids of charges, like in the FQH
effects, or spins. While there is no compelling experimental evidence
of gapped spin liquids on par with that of FQH liquids, there is a
comprehensive body of theoretical work that establishes solvable toy
models where the topological liquid states are apparent. Perhaps some
of the most general, and arguably the most elegant too, are Kitaev's
quantum double models~\cite{kitaev2003fault}. The construction builds
topological states of matter starting from quantum states associated
to elements of a given group. Kitaev's toric code is the simplest such
case, where the group is $\mathbb Z_2$. These constructions, while
exact, contain multi-body interactions; a major open problem is how to
generate these topological states with physical interactions. The
notion of combinatorial gauge symmetry was introduced in
Ref.~\cite{CGY2020} as an effort to address this problem.

Combinatorial gauge symmetry is based on semi-direct or wreath
products of a given symmetry group and permutations, which have
monomial matrix representations with elements in the
group. Hamiltonians with two-body interactions can be constructed so
as to be invariant under a closed string of left and right
multiplications by monomials. These products, along closed paths,
generate an exact local gauge symmetry, and thus these Hamiltonians
contain the same local symmetries as, say, the $\mathbb Z_2$ toric
code.

Examples of systems with $\mathbb Z_2$ combinatorial gauge
symmetry were given for spin systems in Ref.~\cite{CGY2020} and
embedded in a D-Wave quantum annealer in Ref.~\cite{Zhou2020}, and for
superconducting wire arrays in Ref.~\cite{CG2020arXiv}. Here we
provide the first example outside of the family of the simplest type
of $\mathbb Z_2$ topological order, and construct a quantum double for
the group $\mathbb Z_3$ using superconducting wire arrays. This
construction serves as an important stepping stone towards realizing
other quantum doubles within physically accessible Hamiltonians.

The superconducting wire array we present realizes the $\mathbb Z_3$
quantum double on the honeycomb lattice. There are other proposals to
generate $\mathbb Z_n$ quantum doubles with Josephson junction
arrays~\cite{Ioffe2002,Ioffe2003,Ioffe2004}. In those proposals the
gauge symmetry is emergent, i.e., it is realized only in the
perturbative regime where the Josephson energy is dominant. Our
proposal differs in that the gauge symmetry is non-perturbative, i.e.,
the combinatorial gauge symmetry construction discussed here holds for
{\it any} strength of the coupling constants, including regimes where
the charging energy dominates.

The particular construction discussed in this paper has the following
interesting feature: the star potential that usually constrains states
to lie in the zero flux sector is inverted, i.e., the states with
non-zero flux have lowest energy. This inverted potential by itself
would lead to an extensive degeneracy, but the degeneracy can be
lifted by a dimerization pattern encoded in the capacitances of the
array. Of the three wires emanating from a vertex of the honeycomb
lattice, we select one of the directions to have a smaller capacitance
than the other two directions. This choice stabilizes the $\mathbb
Z_3$ topological quantum liquid state. We show that this topological
phase is stable for a range of ratios of the capacitances, up to a
critical ratio for which a quantum XY-ordered phase emerges. We study
the phase diagram and estimate the location of the transition by
deploying a duality map of the model to a quantum three-state Potts
model, which we analyze through a combination of bosonization
techniques (applied to a limit of weakly-coupled one-dimensional
chains) and mean field theory. The dimerization of the couplings
imposed by the different capacitances translate into two different
fields $h_s$ and $h_w$ in the $\mathbb Z_3$ clock model. We estimate
these fields in the effective model in terms of the Josephson energy
and capacitances, using a WKB approximation. We also estimate the size
of the effective plaquette term in the quantum double in terms of
these fields $h_s$ and $h_w$. We point out a positive side-effect of
the inverted potential: the vison gap is larger than that in the case
of the uninverted potential.

The paper is organized as follows. In Sec.~\ref{sec:SCarray} we
present the superconducting wire array that realizes the $\mathbb
Z_3$ combinatorial gauge symmetry yielding the associated topological
quantum double with inverted potential. We show in
Sec.~\ref{sec:topological} that the bond dimerizations, which
microscopically are induced by the different values of the
capacitances in the corresponding wires, lifts the massive degeneracy
imposed by star terms arising from the Josephson couplings and leads
to a $\mathbb Z_3$ topologically ordered ground state. In
Sec.~\ref{sec:quantum-XY} we study the stability of the topological
phase against a quantum XY-ordered state when the degree of
dimerization is reduced. We present a duality transformation into a
$\mathbb Z_3$ clock model. In the appendices we present details of
the calculations, including the estimates of the fields that enter in
the clock model as function of the microscopic parameters of the
superconducting wire array.

\section{Superconducting wire array with combinatorial gauge symmetry}
\label{sec:SCarray}

\begin{figure*}[t]
\centering
\includegraphics[width=.85\textwidth]{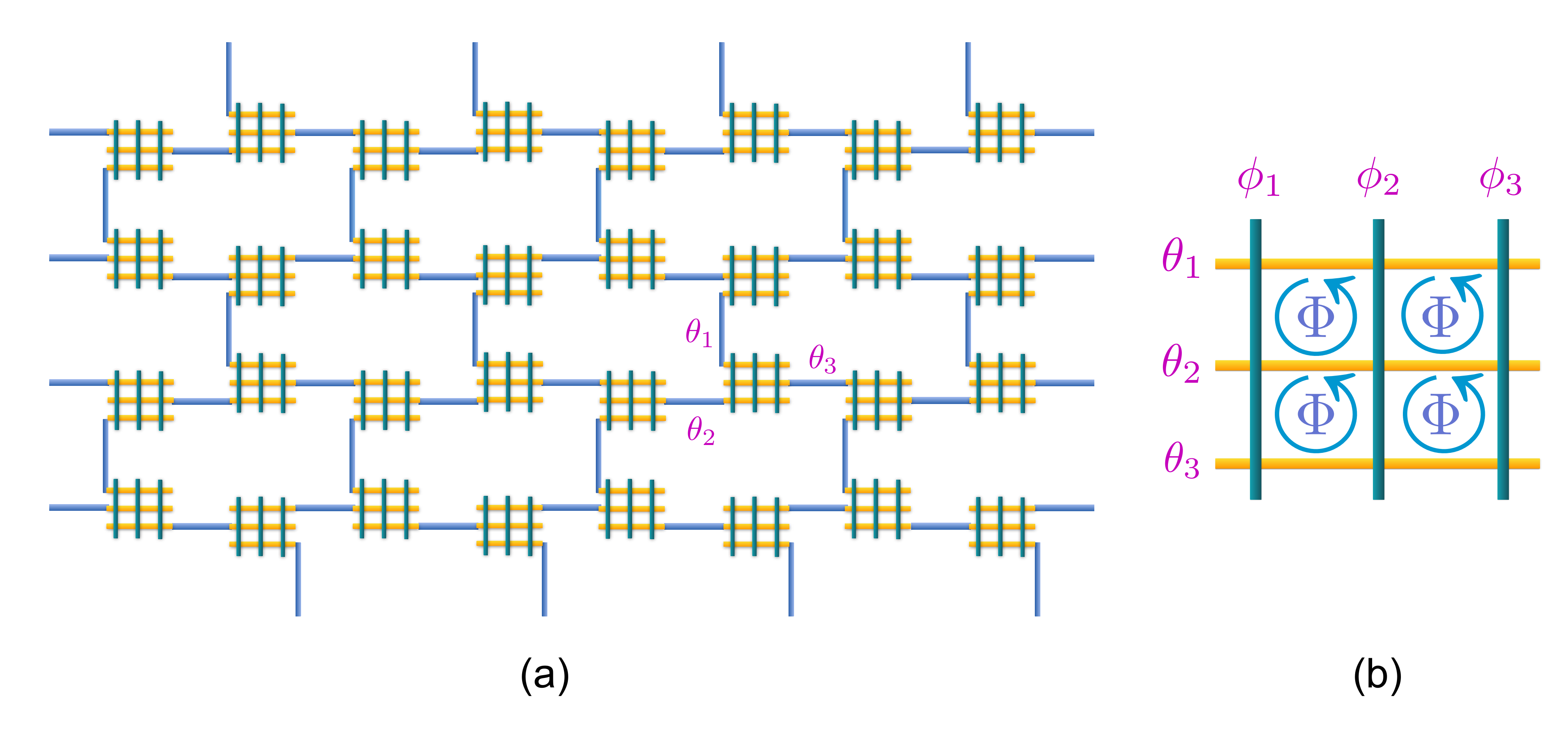}
\caption{An array of superconducting wires forming a two-dimensional honeycomb lattice. An elementary building block is depicted in (b), which contains three horizontal (yellow) gauge wires with superconducting phases $\theta_i$ coupled to three vertical (green) matter wires with phases $\phi_a$ via Josephson junctions, forming a ``waffle'' like geometry. An external magnetic flux of $\Phi = (n+\frac{1}{3})\Phi_0$ threads each elementary plaquette of the waffle, leading to a complex coupling matrix $W$ with combinatorial gauge symmetry. On the full lattice, the gauge wires are shared between two neighboring sites via the blue wires, whereas the matter wires are localized on each lattice site.}
\label{fig:wire}
\end{figure*}

Consider an array of superconducting wires as shown in Fig.~\ref{fig:wire}. An elementary building block depicted in Fig.~\ref{fig:wire}(b) consists of three horizontal ``gauge'' wires and three vertical ``matter'' wires coupled via Josephson junctions, forming a ``waffle'' like geometry. We further introduce an external magnetic flux threading each plaquette of an elementary waffle $\Phi = (n+\frac{1}{3})\Phi_0$, where $n$ is an integer and $\Phi_0 = \frac{h}{2e}$ is the flux quantum.~\footnote{In a more realistic experimental setup, instead of threading flux through the plaquette loops, one can instead replace the single Josephson junction between two crossing wires by ancillary loops forming a highly-asymmetric DC SQUID. The phase shift encoded in the $W$ matrix can be controlled by tuning the flux biases in the two arms of the SQUID. We refer readers to Ref.~\cite{CG2020arXiv} for a more detailed discussion on the experimental perspective.} The full array forms a two-dimensional honeycomb lattice with a waffle at each lattice site, and an extended gauge wire at each link. Notice that the gauge wires are shared between the sites whereas the matter wires are localized on each lattice site.
Denoting the superconducting phases of the gauge wires as $\theta_i$ and the matter wires as $\phi_a$, the Hamiltonian of such an array can be written as
\begin{equation}
H = H_J + H_C,
\label{eq:model}
\end{equation}
where the Josephson coupling
\begin{equation}
H_J = -E_J \sum_s \left[ \sum_{i,a \in s} W_{ai} \ e^{i(\theta_i-\phi_a)} + {\rm h.c.}\right],
\label{eq:H_J}
\end{equation}
and the capacitance term
\begin{equation}
H_C = \frac{1}{2} \sum_s {\bm Q}_{s}^T {\bm C}^{-1} {\bm Q}_{s}.
\label{eq:H_C}
\end{equation}
In the above equations, $E_J$ is the Josephson energy of the
junctions, the vector ${\bm Q}^T = (Q_1, Q_2, Q_3, q_1, q_2, q_3)$
denotes the charge of each gauge wire $Q_i$ and matter wire $q_a$, and
${\bm C}$ is a $6\times 6$ capacitance matrix of the waffle. The
charges and phases are conjugate variables satisfying the standard
commutation relations $[\theta_i, Q_j] = i \delta_{ij}$ and $[\phi_a,
  q_b] = i \delta_{ab}$.

\subsection{Combinatorial gauge symmetry}

The magnetic flux threading each plaquette of the waffle enters the
Josephson coupling as a phase shift, which is encoded in the coupling
matrix $W$ in Eq.~(\ref{eq:H_J}). To see the phase shift in the Josephson coupling energy between each pair of crossing wires $\theta_i$ and $\phi_a$, one simply needs to count the total flux piercing the rectangle formed by wires $(\theta_i, \phi_a, \theta_1, \phi_1)$. For example, consider the
Josephson coupling energy between gauge wire $\theta_2$ and matter
wire $\phi_2$. This coupling acquires a phase shift in the presence of a flux $\Phi$: $-E_J {\rm cos}(\theta_2-\phi_2) \rightarrow -E_J\; {\rm
  cos}(\theta_2-\phi_2 + 2\pi \frac{\Phi}{\Phi_0}) = -E_J\; {\rm
  cos}(\theta_2-\phi_2 + \frac{2\pi}{3})$. This corresponds to matrix element $W_{22} = e^{i\frac{2\pi}{3}}$ in Eq.~(\ref{eq:H_J}).
Similarly, the phase shift between wire $\theta_1$ and any matter wire $\phi_a$, as well as wire $\phi_1$ and any gauge wire $\theta_i$, is zero, since there is no loop formed in this case, which corresponds to $W_{1i}=W_{a1}=1$.
One can readily check that the coupling
matrix $W$ in Eq.~(\ref{eq:H_J}) has the following form:
\begin{equation}
W = \frac{1}{\sqrt{3}}
\begin{pmatrix}
1 & 1 & 1 \\
1 & \omega  & \overline{\omega}  \\
1 & \overline{\omega}  & \omega
\end{pmatrix},
\label{eq:W}
\end{equation}
where $\omega = e^{i\frac{2\pi}{3}}$ and $\overline{\omega} =
\omega^2$. One recognizes that the $W$ matrix above is precisely the
discrete Fourier transform matrix with entries $W_{jk} =
\frac{1}{\sqrt{3}} e^{i \frac{2\pi (j-1)(k-1)}{3}}$, which is also a
complex Hadamard matrix satisfying $W^\dagger W = W
W^\dagger=\mathbb{1}$. Complex Hadamard matrices of the
form~(\ref{eq:W}) are invariant under a pair of left/right monominal
transformations, which underlie the combinatorial gauge
symmetry. Specifically, $W$ has the following automorphism
\begin{equation}
L^\dagger \ W \ R = W,
\label{eq:auto}
\end{equation}
where $L$ and $R$ are monomial matrices. Equivalently, $R$ and $L$ generate permutations and shifts of the superconducting phases on the gauge and matter wires within a waffle, respectively:
\begin{subequations}
\begin{align}
e^{i \theta_i} &\rightarrow  \sum_{j=1}^3 R_{ij} \; e^{i\theta_j}, \\
e^{-i \phi_a}  &\rightarrow  \sum_{b=1}^3 e^{-i \phi_b}\; (L^\dagger)_{ba},
\label{eq:L}
\end{align}
\label{eq:RL}
\end{subequations}
under which the Josephson coupling terms in the Hamiltonian of a
single waffle is invariant. We further restrict the $R$ matrix to be
diagonal, since the gauge wires on the lattice are shared between
sites and cannot be permuted. It turns out that if we take $R$ to be
of the following form
\begin{equation}
R =
\begin{pmatrix}
1 & 0 & 0 \\
0 & \omega  & 0  \\
0 & 0 & \overline{\omega}
\end{pmatrix},
\end{equation}
then $L$ is also a monomial matrix
\begin{equation}
L =
\begin{pmatrix}
0 & 1 & 0 \\
0 & 0 & 1 \\
1 & 0 & 0
\end{pmatrix}.
\end{equation}
One can further check that $L$ is monomial for any permutation along the diagonal of the $R$ matrix. Notice that $L$ is uniquely determined for a given $R$, following from the automorphism Eq.~(\ref{eq:auto}). Permutations of the matter wires are allowed on the lattice because the matter wires are localized on each site, and permutations simply correspond to relabeling the wires.
Physically, the transformation $R$ corresponds to a $\mathbb{Z}_3$ phase shift on two out of the three gauge wires within a waffle, such that the product $\prod_{i=1}^3 e^{i\theta_i}$ is preserved.

\begin{figure}[t]
\centering
\includegraphics[width=.38\textwidth]{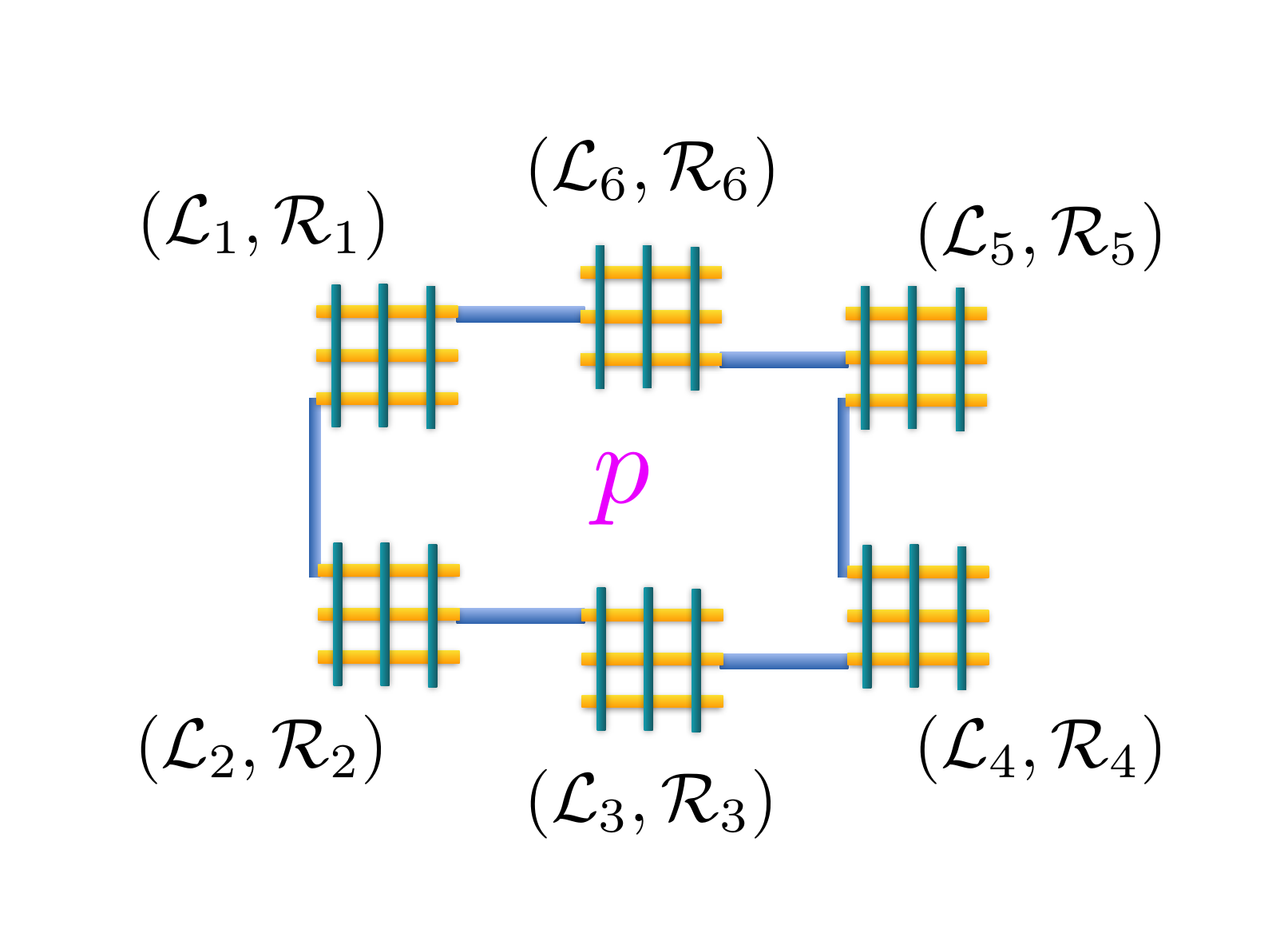}
\caption{Local $\mathbb{Z}_3$ gauge symmetry on an elementary hexagonal plaquette of the lattice.}
\label{fig:gauge_symmetry}
\end{figure}

The automorphism of $W$ under monomial transformations $(L, R)$ naturally furnishes a local $\mathbb{Z}_3$ gauge symmetry on the full lattice. One can construct the following operator generating a local gauge transformation around an elementary hexagonal plaquette on the lattice, as depicted in Fig.~\ref{fig:gauge_symmetry}:
\begin{equation}
G_p = \prod_{s \in p} \mathcal{L}_s^{(\phi)} \prod_{s \in p} \mathcal{R}_s^{(\theta)},
\label{eq:gauge_generator}
\end{equation}
where $\mathcal{L}_s^{(\phi)}$ generates permutations and phase shifts of the matter wires located on site $s$ according to Eq.~(\ref{eq:L}): $\mathcal{L}_s^{(\phi)} e^{-i\phi_a} (\mathcal{L}_s^{(\phi)})^{-1} = \sum_{b} e^{-i \phi_b} (L^\dagger)_{ba}$, and $\mathcal{R}_s^{(\theta)}$ generates phase shifts on the two gauge wires emanating from site $s$: $\mathcal{R}_s^{(\theta)} e^{i\theta_i} (\mathcal{R}_s^{(\theta)})^{-1} =  \sum_{j} R_{ij} e^{i\theta_j}$. The automorphism of $W$ directly leads to the invariance of Hamiltonian~(\ref{eq:H_J}) under $G_p$: $[G_p, H_J]=0$, for all $p$. Furthermore, the local gauge transformations on different plaquettes commute with one another: $[G_p, G_{p'}]=0$. Thus we have shown that the system in the classical limit where only $H_J$ is present has a local $\mathbb{Z}_3$ gauge symmetry.  Next, we will show that the capacitance term $H_C$ is also invariant under $G_p$.

The capacitance matrix ${\bm C}$ contains the following entries: the
self-capacitances of a single gauge wire, $C_g$, and of a single
matter wire, $C_m$; the capacitance of the Josephson junction, $C_J$;
and the mutual-capacitance between two neighboring wires that are
parallel to one another, $C_p$. The capacitance $C_p$ is the smallest
of all, as can be easily inferred from the geometry if the wires are
thin and widely separated compared to their width. Neglecting $C_p$
yields a capacitance matrix ${\bm C}$ that is invariant under the
permutation of the indices of the matter (as well as gauge)
wires. This symmetry carries to the inverse matrix ${\bm C}^{-1}$ that
controls the charging energies.~\footnote{We remark that a small value
  of $C_p$ breaks the permutation symmetry among the three matter
  wires; nevertheless, if Hamiltonian~(\ref{eq:model}) supports a
  gapped phase with $\mathbb{Z}_3$ topological order, it will remain
  stable in the presence of a small combinatorial symmetry breaking
  perturbation so long as the gap stays open. In practice, one can also design the geometry of the wires, such that the $C_p$'s between different wires are symmetrized. See Ref.~\cite{CG2020arXiv} for details.}

Since the charge and superconducting phases are conjugate variables,
the $\mathbb{Z}_3$ phase shift in the monomial transformations $(L,
R)$ is generated by the unitary operators
\begin{equation}
U^{(R)}_i = e^{\pm i \frac{2\pi}{3} Q_i}, \quad  U^{(L)}_a = e^{\pm i \frac{2\pi}{3} q_a}
\end{equation}
acting on the gauge and matter wires, respectively. Because these
unitary operators trivially commute with the charge operators $Q_i$
and $q_a$, and ${\bm C}^{-1}$ is invariant under the permutation of
the matter wires, we conclude that $[G_p, H_C]=0$. Combining with the
previous finding that $[G_p, H_J]=0$, it follows that $[G_p,
  H]=0$. Hence the full lattice Hamiltonian~(\ref{eq:model}) is a
gauge theory with local $\mathbb{Z}_3$ combinatorial gauge symmetry.

\subsection{Minima of $H_J$ on a single waffle}
\label{sec:minima}

\begin{figure*}[t]
\centering
\includegraphics[width=.8\textwidth]{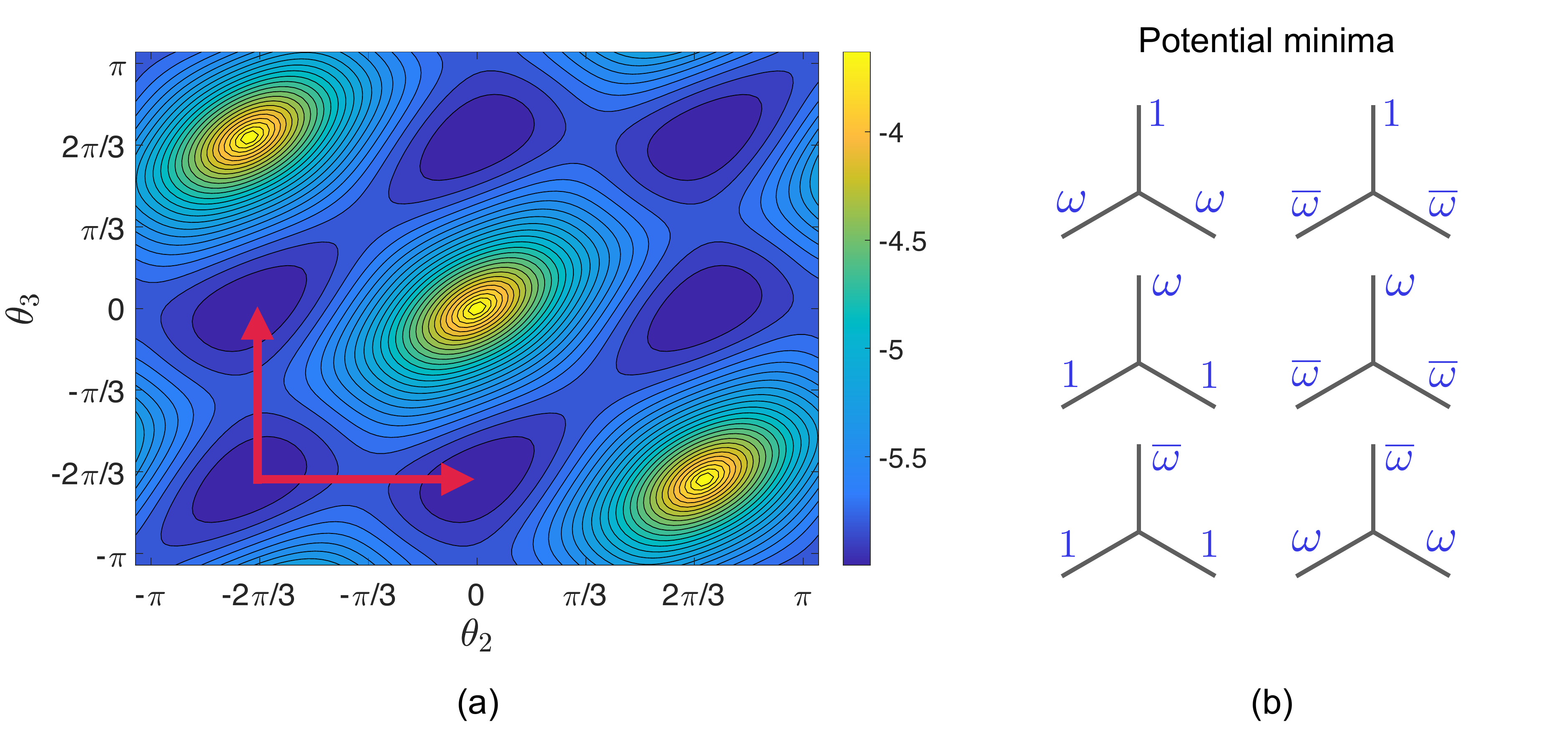}
\caption{(a) Contour plot of the potential $-2 \sum_a \left|\sum_i W_{ai} e^{i\theta_i} \right|$ as a function of $\theta_2$ and $\theta_3$. We fix $\theta_1=0$. (b) All inequivalent gauge wire phase configurations $(\theta_1, \theta_2, \theta_3)$ at the potential minima. For each configuration, there are another two equivalent ones from permuting the three phases, yielding a total of 18 configurations. The ground state configurations can be represented by a $\mathbb{Z}_3$ variable $Z$, such that at each star $\prod_{i \in s} Z_i = \omega$ or $\overline{\omega}$. Configurations satisfying $\prod_{i \in s} Z_i = 1$ correspond to the maxima of the potential.
The capacitance term generates tunneling processes (instanton) between nearest-neighboring minima, which corresponds to a transverse field term in the effective $\mathbb{Z}_3$ representation: $X+X^\dagger$. The red arrows in (a) show two examples of such tunneling processes.}
\label{fig:potential}
\end{figure*}

Having established the local combinatorial gauge symmetry of
Hamiltonian~(\ref{eq:model}), we shall now look at the minima of the
classical potential energy $H_J$ on a single waffle. We will show that
the superconducting phases of the gauge wires at the potential minima
are $\mathbb{Z}_3$-valued. Therefore, when the Josephson energy is the
dominant scale, Hamiltonian~(\ref{eq:model}) can be effectively
described in terms of $\mathbb{Z}_3$ variables, from which the
topological phase emerges.

Minimizing the Josephson energy ties together the $\phi_a$ and
$\theta_i$ phase variables (see details in Appendix~\ref{sec:app:phases}),
\begin{equation}
e^{i\phi_a} = \frac{\sum_i W_{ai} e^{i\theta_i}}{\left|\sum_i W_{ai} e^{i\theta_i}\right|}.
\label{eq:phi}
\end{equation}
The minimum energy is given by
\begin{equation}
E_{\rm min}
  =
-2E_J \sum_a \left|\sum_i W_{ai}
e^{i\theta_i} \right|
\;.
\label{eq:minE}
\end{equation}
We plot the potential profile as a function of $\theta_2$ and
$\theta_3$ while fixing $\theta_1=0$ in
Fig.~\ref{fig:potential}(a). We find six (three inequivalent)
degenerate minima with $E_{\rm min} =-6E_J$ corresponding to
$\theta_2$ and $\theta_3$ being 0 or $\pm \frac{2\pi}{3}$, such that
$\prod_i e^{i\theta_i}= \omega$ or $\overline{\omega}$. In
Fig.~\ref{fig:potential}(b), we show all six inequivalent gauge wire
phase configurations corresponding to the minima of the potential
energy. For each configuration shown in Fig.~\ref{fig:potential}(b),
there are another two equivalent ones from permuting the three phases,
yielding a total of 18 ground state configurations. Notice that the
three degenerate \textit{maxima} in Fig.~\ref{fig:potential}(a) with
$E_{\rm max}= -2\sqrt{3}E_J$ also correspond to $\theta$ being 0 or
$\pm \frac{2\pi}{3}$, but now with $\prod_i e^{i\theta_i}=1$ instead.

Turning on the capacitance term $H_C$ introduces quantum fluctuations
in the phases. When the Josephson energies are larger than the
charging energies, $H_C$ induces tunneling between nearest-neighboring
minima, which corresponds to an instanton in Euclidean spacetime. In
Fig.~\ref{fig:potential}(a), we show two examples of such tunneling
processes at leading order, where one of the three phases is shifted
by $\pm \frac{2\pi}{3}$ while the other two remain
unchanged. Semiclassically, the amplitude of such a tunneling process
can be estimated from the Euclidean action of the instanton (or
equivalently, the WKB approximation). We provide detailed calculations
in Appendix~\ref{app:instanton}, which lead to a tunneling amplitude
$\sim \exp\left({-0.88\sqrt{C_{\rm eff} E_J}}\right)$, where $C_{\rm eff}$ is an
effective capacitance dependent on $C_g, C_m$, and $C_J$. Notice that
this amplitude already takes into account the shifts in $\phi_a$,
which, when $E_J$ is large, follow the instantaneous minimum of $H_J$
and is hence locked to $\theta_i$ according to Eq.~(\ref{eq:phi}).

\begin{figure}[t]
\centering
\includegraphics[width=.38\textwidth]{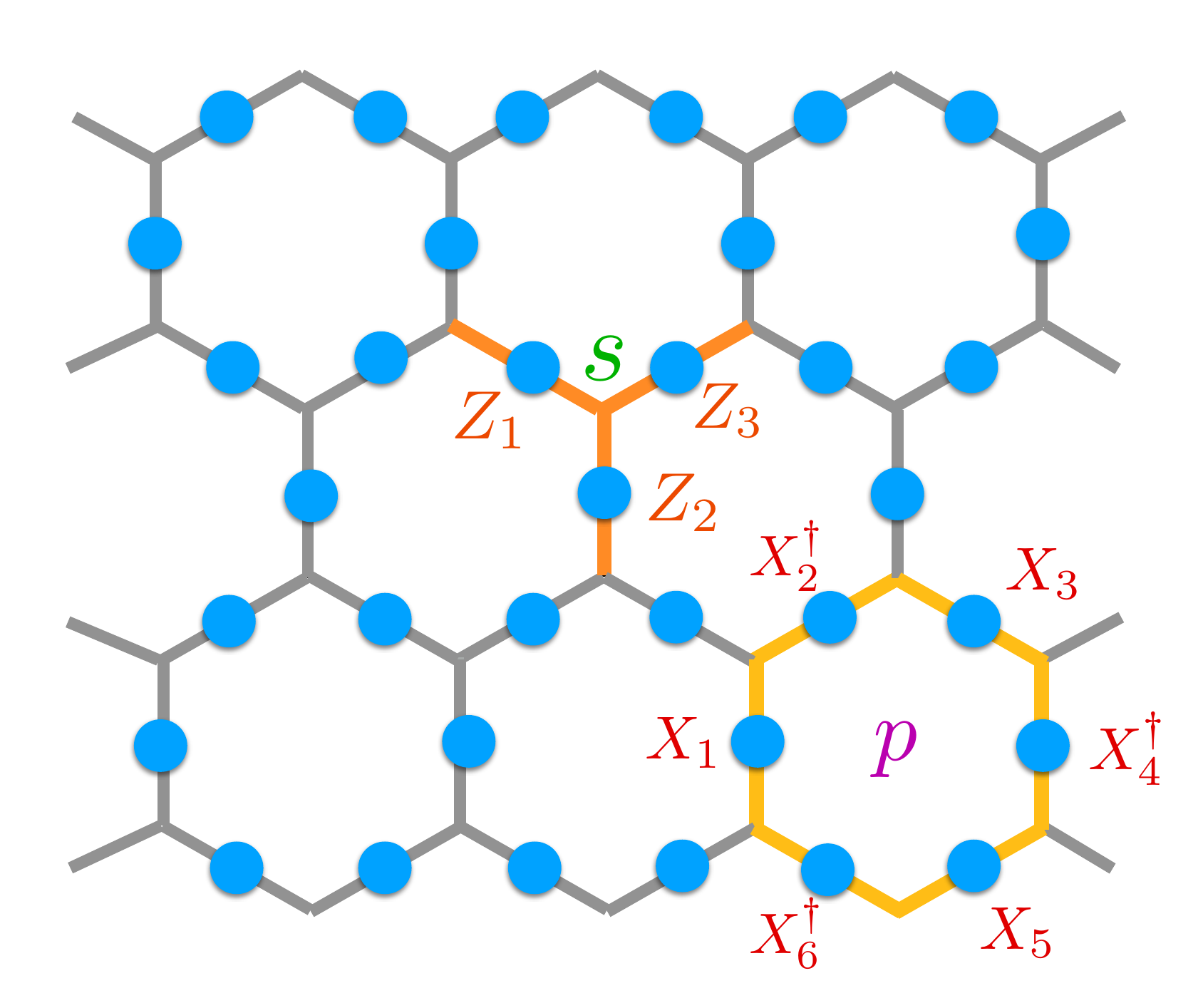}
\caption{Low energy description of the superconducting wire array in term of $\mathbb{Z}_3$ clock variables (blue dots) on the link of a honeycomb lattice. The star operator $A_s$ and the generator of local $\mathbb{Z}_3$ gauge transformation $G_p$ are highlighted.}
\label{fig:clock}
\end{figure}

The minima depicted in Fig.~\ref{fig:potential} suggest that the system admits an effective representation in terms of $\mathbb{Z}_3$-valued operators at low energy. We place a $\mathbb{Z}_3$ degree of freedom on each bond of the honeycomb lattice (Fig.~\ref{fig:clock}), and introduce $\mathbb{Z}_3$ clock operators $Z_i$ and $X_i$ satisfying the algebra
\begin{subequations}
\begin{align}
X_i^3 = Z_i^3=1, \quad X_i^\dagger = X_i^2,  \quad Z_i^\dagger = Z_i^2,  \\
Z_i X_j = \omega^{\delta_{ij}} X_j Z_i, \quad  Z_i X_j^\dagger = \overline{\omega}^{\delta_{ij}} X_j^\dagger Z_i.
\end{align}
\end{subequations}
In the basis where $Z$ is diagonal, $Z=\{1, \ \omega,\ \overline{\omega}\}$ represents the three possible gauge wire phases $e^{i\theta_i}$ at the potential minima. The capacitance-induced tunneling can be represented by a transverse field $X+X^\dagger$ that shifts the $Z$ eigenvalue by $\pm \frac{2\pi}{3}$. In terms of the clock variables, the superconducting wire array can be described effectively at low energy by
\begin{equation}
H_{\mathbb{Z}_3} = J \sum_s (A_s + A_s^\dagger) - h \sum_i (X_i + X_i^\dagger),
\label{eq:clock}
\end{equation}
where $A_s = \prod_{i \in s} Z_i$, $J=\frac{6-2\sqrt{3}}{3}E_J>0$, and $h \sim e^{-0.88\sqrt{C_{\rm eff} E_J}}$. The generator of local $\mathbb{Z}_3$ gauge transformation~(\ref{eq:gauge_generator}) now takes the form $G_p = X_1 X_2^\dagger X_3 X_4^\dagger X_5 X_6^\dagger$ and $G_p^\dagger$ around a hexagonal plaquette, as shown in Fig.~\ref{fig:clock}. It is easy to verify that $[G_p, A_s]=[G_p, A_s^\dagger]=0$ for any $p$, $s$; hence $[G_p, H_{\mathbb{Z}_3}]=0$. With periodic boundary conditions, $A_s$ and $G_p$ satisfy the following constraints:
\begin{equation}
\prod_p G_p = 1,  \quad  \prod_{s \in A} A_s \prod_{s \in B} A_s^\dagger = 1,
\label{eq:constraint}
\end{equation}
where $A$ and $B$ denote two sublattices of the honeycomb lattice.  In
Appendix~\ref{app:equivalence}, we show explicitly that
Hamiltonian~(\ref{eq:clock}) with the gauge constraint imposed by
$G_p$ is equivalent to the $\mathbb{Z}_3$ quantum double
model~\cite{kitaev2003fault}. However, since $J>0$, the star term in
Hamiltonian~(\ref{eq:clock}) energetically favors $A_s=\omega$ or
$\overline{\omega}$, while $A_s=1$ has a higher energy. In other
words, we are sitting within a non-zero mixed flux sector of the
quantum double model due to the inverted potential. As we will see in
the next section, this key distinction from the conventional quantum
double model has important consequences on the phase diagram of the
system, and in particular, on how the topologically ordered phase
emerges. In Appendix~\ref{sec:app:uninverted}, we also provide a $W$
matrix leading to the usual $\mathbb{Z}_3$ quantum double where the
star term favors the zero-flux sector $A_s=1$.

\section{$\mathbb{Z}_3$ topologically ordered phase from bond dimerization}
\label{sec:topological}
\begin{figure*}[t]
\centering
\includegraphics[width=.85\textwidth]{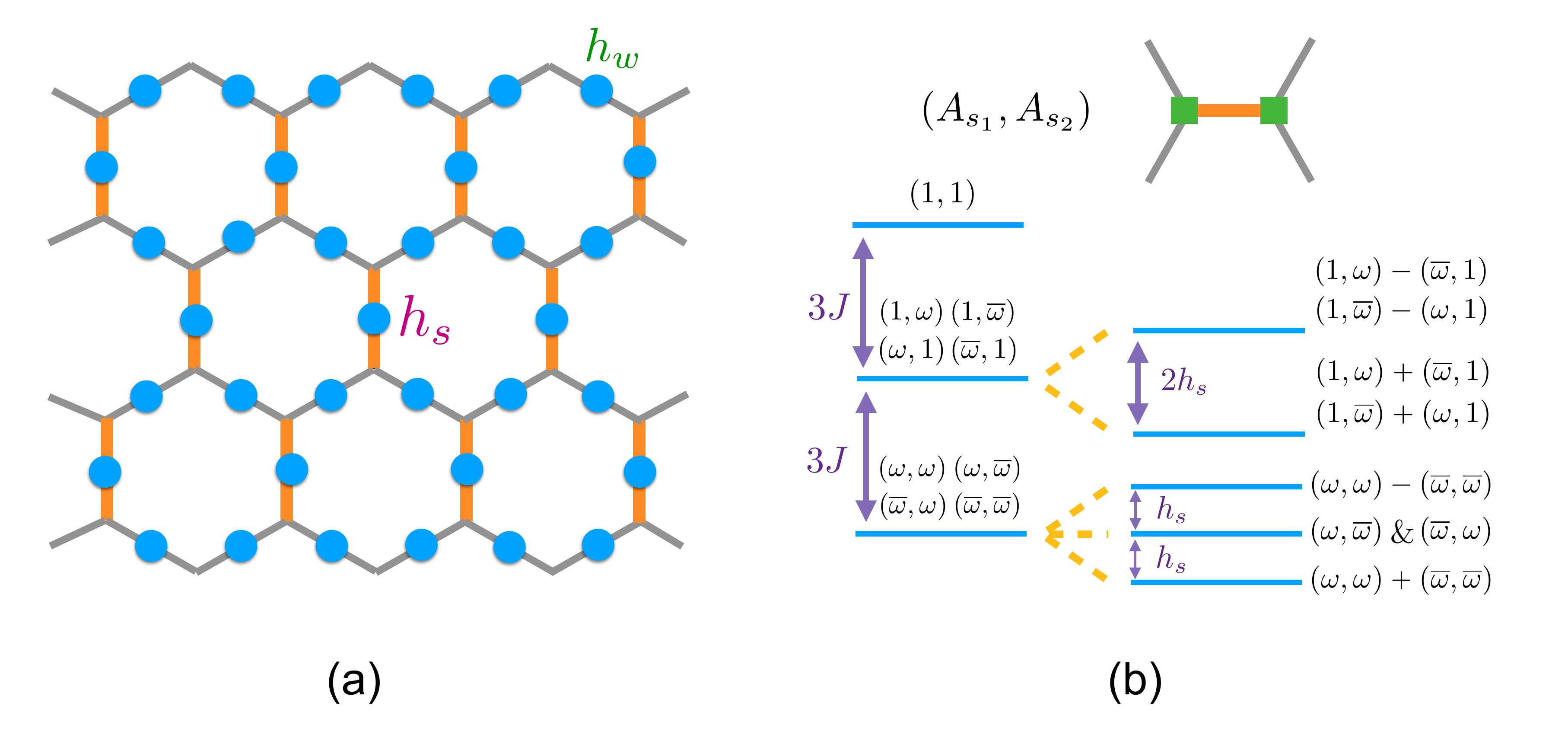}
\caption{(a) A strong transverse field $h_s$ is applied on all vertical (orange) bonds, which form the hexagonal lattice; and a weak transverse field $h_w$ is applied on all other (grey) bonds. (b) Energy levels of a single dimer formed by the strong bond in the limit of infinite $J$ and $h_w=0$. In the absence of $h_s$, the ground state is four-fold degenerate in $(A_{s_1}, A_{s_2})$, which is the source of the massive ground state degeneracy in Eq.~(\ref{eq:gsd1}). A non-zero $h_s$ splits the four-fold degeneracy, leading to a unique ground state in terms of the star variables. This leads to the nine-fold topological ground state degeneracy in Eq.~(\ref{eq:gsd2}) on the full lattice.}
\label{fig:dimerization}
\end{figure*}

We shall now discuss the phases that Hamiltonian~(\ref{eq:clock})
sustains. For the conventional quantum double model with a $-J$ in
front of the star term, one expects a gapped phase with $\mathbb{Z}_3$
topological order for $h/J < (h/J)_c$. However, the situation is
drastically different for an inverted potential with a $+J$ in front
as in Hamiltonian~(\ref{eq:clock}). Let us start by counting the
ground state degeneracy in the limit $J \rightarrow \infty$ (or
$h=0$), and when the gauge constraint $G_p=G_p^\dagger=1$ is
imposed. Denote the total number of vertices, bonds, and plaquettes on
the honeycomb lattice as $N_v$, $N_b$, and $N_p$, respectively. The
ground state degeneracy of Hamiltonian~(\ref{eq:clock}) on a torus in
the large $J$ limit is
\begin{equation}
{\rm GSD} = 3^{N_b} \times \left(\frac{1}{3}\right)^{N_v-1}\times 2^{N_v} \times \left( \frac{1}{3}\right)^{N_p-1} = 2^{N_v}\times 3^2,
\label{eq:gsd1}
\end{equation}
where we have used the relations $N_b=3N_p=3N_v/2$, and the -1's on
the exponents account for the constraint~(\ref{eq:constraint}). This
indicates that the ground state is massively degenerate, and that the
gauge constraint cannot fully lift this degeneracy. We will show in
the next section that upon further turning on a weak uniform
transverse field $h$, the system can be mapped to a quantum
spin-$\frac{1}{2}$ XY model which is in fact gapless. Therefore, due
to the inverted potential, Hamiltonian~(\ref{eq:clock}) as it is does
not support a gapped topological phase.

Nevertheless, a gapped topological phase emerges with a slight
modification of Hamiltonian~(\ref{eq:clock}). Instead of a uniform
transverse field, we apply a strong transverse field $h_s$ on the
vertical bonds forming the hexagonal lattice, and a weak transverse
field $h_w$ on all other bonds with $h_s>h_w$ while keeping both $h_s$
and $h_w$ much smaller than $J$. (Notice that the WKB calculation yields an exponential suppression in the tunneling amplitudes $h_s$ and $h_w$. Thus it is experimentally feasible to have $h_s$ and $h_w$ smaller than $J$~\cite{CG2020arXiv}.) This leads to a bond dimerization
pattern depicted in Fig.~\ref{fig:dimerization}(a). Now the
Hamiltonian takes the form
\begin{eqnarray}
H_{\mathbb{Z}_3} = &&J \sum_s (A_s + A_s^\dagger) - h_s \sum_{i \in {\rm vertical}} (X_i + X_i^\dagger)  \nonumber \\
&&- h_w \sum_{i \notin {\rm vertical}} (X_i+X_i^\dagger).
\label{eq:dimerization}
\end{eqnarray}
We start by considering the limit when $h_w=0$. In this limit, the
system becomes a set of decoupled dimers formed by the strong bonds,
since the weak bonds have no dynamics. In
Fig.~\ref{fig:dimerization}(b), we show the energy levels associated
with a single dimer in the limit of infinite $J$. In the absence of
$h_s$, the ground state of the single dimer is four-fold degenerate in
$(A_{s_1}, A_{s_2})$ corresponding to each $A_s=\omega$ or
$\overline{\omega}$, which is the source of the massive ground state
degeneracy in Eq.~(\ref{eq:gsd1}).  The excited states correspond to
flipping either or both stars to $A_s=1$, which is separated from the
ground state subspace by a large energy of order $J$. Upon turning on
$h_s$, the four degenerate ground states will split, and the unique
ground state, in the star variables, is $\frac{1}{2}(|\omega \omega
\rangle + |\overline{\omega} \overline{\omega} \rangle)$, whose energy
is lowered by $h_s$. To show that the massive degeneracy on the full
lattice is indeed split, we compute the ground state degeneracy in
this case:
\begin{equation}
{\rm GSD} = 3^{N_b} \times \left(\frac{1}{3}\right)^{N_v-1}\times 2^{N_v} \times \left( \frac{1}{3}\right)^{N_p-1} \times \left(\frac{1}{4} \right)^{N_p} = 3^2,
\label{eq:gsd2}
\end{equation}
where the additional factor of $\left(\frac{1}{4}\right)^{N_p}$
corresponds to one constraint per unit cell imposed by $h_s$, and the
total number of unit cells is equal to $N_p$. We find that the ground
state indeed has a nine-fold topological degeneracy on a torus, which
coincides with that of the conventional $\mathbb{Z}_3$ quantum double.

In the above countings, the gauge constraint $G_p=G_p^\dagger=1$ is
imposed by hand. In our model, such a plaquette term can be generated
perturbatively upon turning on $h_w$, yielding an associated energy
scale corresponding to the vison gap. Here we point out another key
distinction from the usual $\mathbb{Z}_3$ quantum double with a $-J$ in
the Hamiltonian. In that case, a plaquette term is generated only at
sixth order in $h/J$ in degenerate perturbation theory, yielding a
very small vison gap when $h/J < (h/J)_c$. The reason for such a small
vison gap is that $h$ creates star excitations with a large energy
cost of order $J$, which suppresses the gap. In our
model~(\ref{eq:dimerization}) with an inverted potential and dimerized
transverse fields, however, one does not have to pay an energy of
order $J$ to create an excitation. Rather, there are cheaper
excitations one can make that only cost an energy of order $h_s$,
which correspond to transitioning between the ground state and first
excited state in the presence of $h_s$ as shown in
Fig.~\ref{fig:dimerization}(b). Furthermore, the plaquette term now
can be generated at fourth order in $h_w/h_s$, leading to a larger
vison gap than in the usual $\mathbb{Z}_3$ quantum double.~\footnote{We remark that because the resulting vison gap is small compared to the scale $J$, it
would be difficult in practice to attain low enough temperatures to
reach the true ground state, or even a thermal state with a low
density of visons. Nonetheless, there may still be signatures of the
mutual statistics of the spinons and visons that could be observed in
the regime where temperature is larger than the vison gap but still
much smaller than the spinon gap, as discussed (for the $\mathbb Z_2$
model) in Ref.~\cite{hart2021correlation}.}

\begin{figure}[t]
\centering
\includegraphics[width=.48\textwidth]{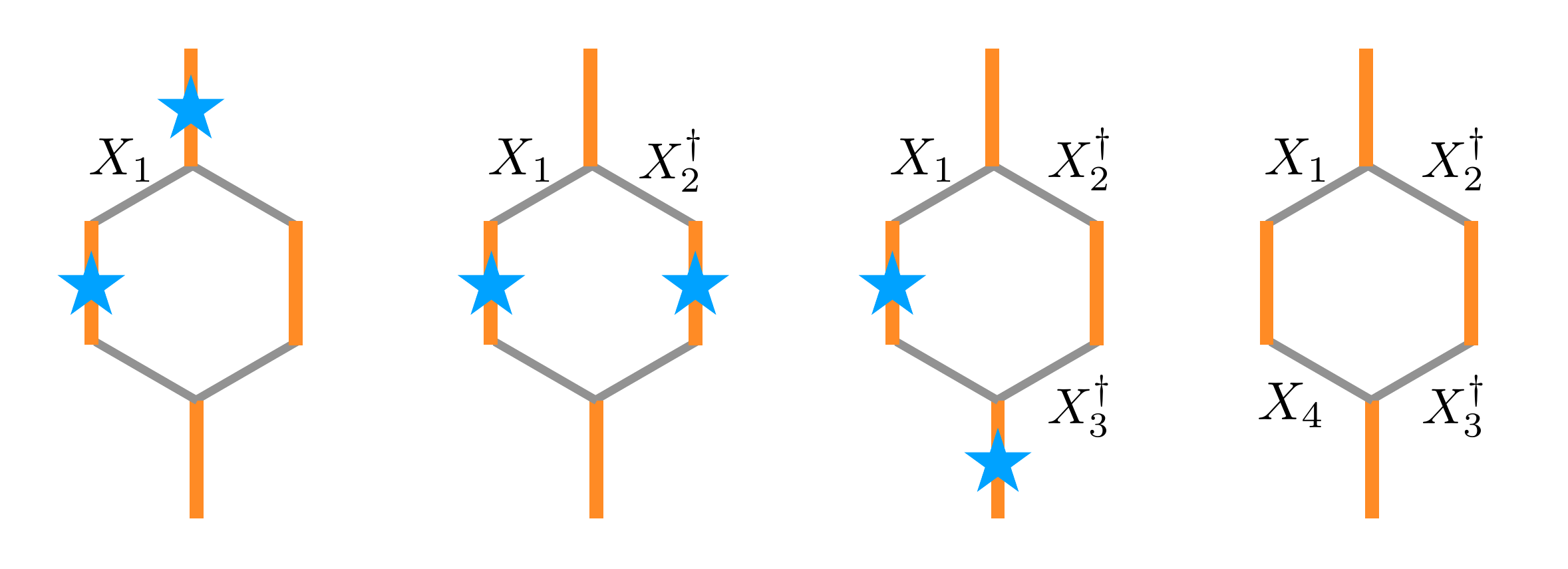}
\caption{A fourth-order process in degenerate perturbation theory where a pair of excitations are created and annihilated around a plaquette under the action of $h_w$.}
\label{fig:perturbation}
\end{figure}

We compute the plaquette term from Hamiltonian~(\ref{eq:dimerization}) using degenerate perturbation theory. In Fig.~\ref{fig:perturbation} we show a fourth-order process in which a pair of excitations is created and annihilated around a plaquette. In the absence of $h_w$, the ground state written in terms of star variables is a tensor product of $\frac{1}{\sqrt{2}}(|\omega \omega\rangle + |\overline{\omega} \overline{\omega}\rangle)$ on all strong bonds. Applying $h_w$ on a weak bond shifts the two stars connected to the weak bond by $\omega$ or $\overline{\omega}$. For example, consider the action of a weak transverse field on a bond connected to $s_1$ of a dimer $(s_1, s_2)$ in its ground state:
\begin{eqnarray}
&&-h_w (X_1 + X_1^\dagger)\ \frac{1}{\sqrt{2}} \left(|\omega \omega\rangle + |\overline{\omega}\overline{\omega}\rangle\right)  \nonumber  \\
&&\rightarrow |\overline{\omega}\omega\rangle  + |\omega\overline{\omega}\rangle + |1\overline{\omega}\rangle + |1\omega\rangle,
\end{eqnarray}
where the last two states cost an energy of order $J$ and can be
projected out in the limit of infinite $J$. The first two states, on
the other hand, are low energy excitations with energy $h_s$
only. Define the projector onto the low energy subspace:
  $Q = |\overline{\omega} \omega \rangle \langle
  \overline{\omega}\omega| + |\omega \overline{\omega} \rangle \langle
  \omega \overline{\omega}|$.  One can thus compute the effective
Hamiltonian at fourth order in $h_w/h_s$ acting within the ground
state subspace. For example, the process shown in
  Fig.~\ref{fig:perturbation} gives a contribution to the effective
  Hamiltonian:
\begin{equation}
  H_{\rm eff} \supset -\frac{h_w^4}{(2h_s)^3} \sum_p \left[ X_1  Q X_2^\dagger Q X_3^\dagger Q X_4 + {\rm h.c.} \right]
  \label{eq:supset}
\end{equation} 
There are in total 24 different fourth-order processes of pair
creations and annihilations around a plaquette, and contributions from
all other processes can be calculated in a straightforward manner. The
important point here is that the perturbative vison gap is fourth
order in $h_w/h_s$ as opposed to sixth order in $h/J$, and hence can
be made significantly larger than that in the usual $\mathbb{Z}_3$
quantum double. In Appendix~\ref{app:spider}, we provide numerical
results of Hamiltonian~(\ref{eq:dimerization}) on an elementary
``spider'' like geometry, and compare with the conventional
$\mathbb{Z}_3$ quantum double with $-J$. The numerical results indeed
suggest that a larger gap can be achieved in our model.

To conclude, Hamiltonian~(\ref{eq:dimerization}) sustains a gapped
phase with $\mathbb{Z}_3$ topological order upon introducing strong
and weak transverse fields as depicted in
Fig.~\ref{fig:dimerization}. Our analysis above mainly focuses on the
perturbative regime where $J \gg h_s \gg h_w$, but we expect the
topological phase to persist for $h_s/J < (h_s/J)_c$, and
$h_w<h_s$. For $h_s/J > (h_s/J)_c$, the transverse field dominates and
the system becomes a trivial paramagnet. In the next section, we will
consider the regime where $h_w \geq h_s$, and $h_s/J < (h_s/J)_c$.

\section{Quantum XY-ordered phase}
\label{sec:quantum-XY}

Now that we have established the existence of a topological phase in our system, let us now consider what happens if $h_w$ becomes greater than $h_s$ while both $h_s/J$ and $h_w/J$ are small. In this regime, since the star operator still has a non-zero expectation value in the ground state, it is useful to consider a dual description of Hamiltonian~(\ref{eq:dimerization}) in terms of $\mathbb{Z}_3$ clock degrees of freedom on the vertices of the honeycomb lattice, which we have been implicitly using in the previous section. In this section, we shall first show a duality mapping from Hamiltonian~(\ref{eq:dimerization}) to a quantum three-state Potts model. In the dual picture, the isotropic point $h_w=h_s$ maps to a quantum spin-$\frac{1}{2}$ XY model with XY ordering in the ground state and a gapless spectrum~\cite{PhysRevLett.61.2582, PhysRevB.60.6588}. For $h_w>h_s$, we consider the limit $h_s=0$, when the system maps to decoupled XY chains. A small $h_s$ couples the chains, and we analyze the effect of inter-chain couplings using abelian bosonization. We find that the inter-chain coupling due to a weak $h_s$ is marginal around the decoupled chain fixed point, and hence the system should remain gapless for a non-zero but weak $h_s$.

\begin{figure}[t]
\centering
\includegraphics[width=.45\textwidth]{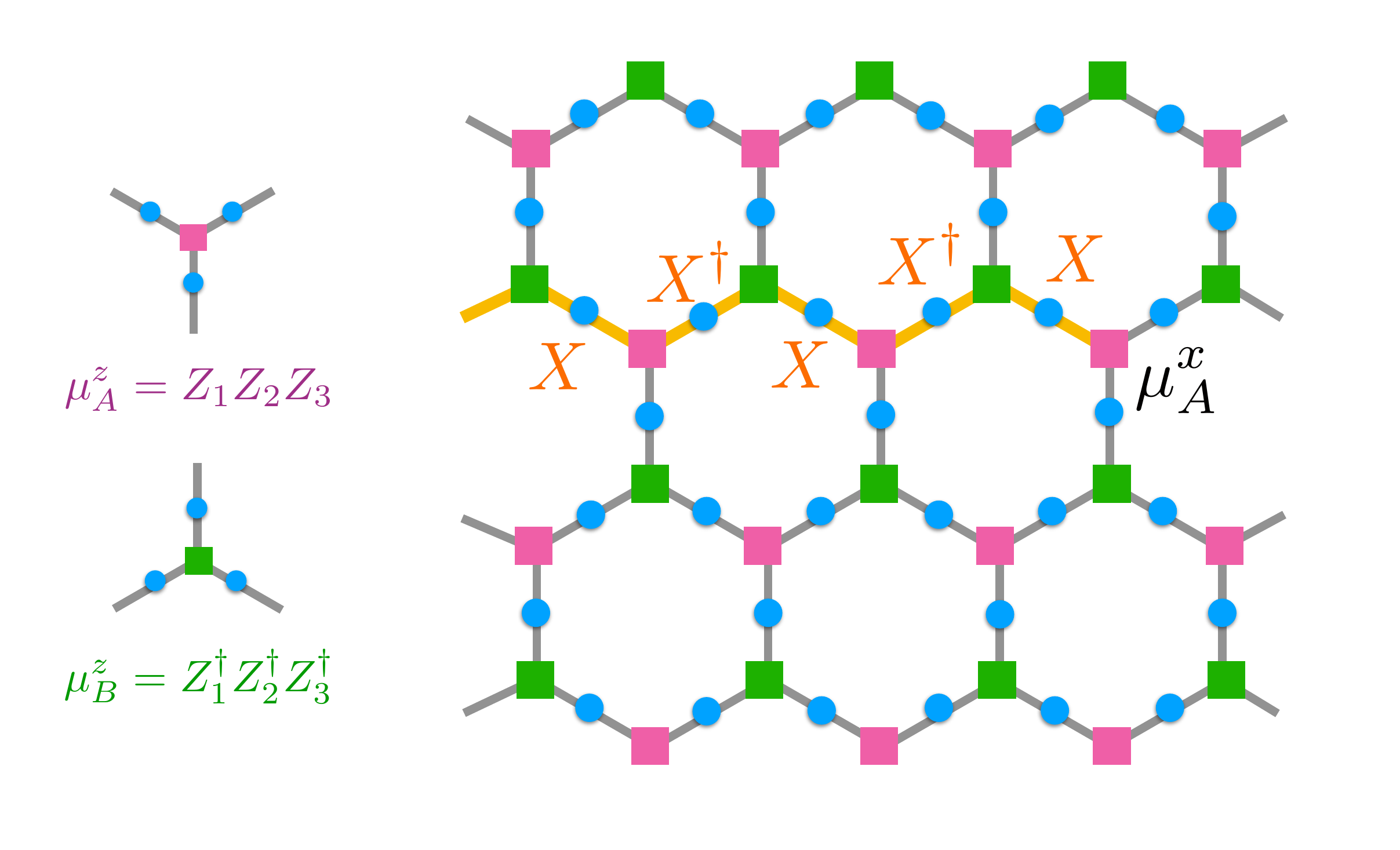}
\caption{Duality transformations defined in Eqs.~(\ref{eq:duality}) and~(\ref{eq:string}). Gauge degrees of freedom in the original model are placed on the links (blue dots), and the dual clock variables are placed on the vertices (squares). The mappings for $\mu^z$ are different for sublattice $A$ (purple squares) and $B$ (green squares). A single $\mu^x$ operator is expressed as a string operator in terms of the gauge degrees of freedom.}
\label{fig:duality}
\end{figure}

\subsection{Duality mapping: quantum three-state Potts model}

The duality we demonstrate here is in close analogy with the familiar Kramers-Wannier duality between the two-dimensional transverse field Ising model and the $\mathbb{Z}_2$ quantum double~\cite{wegner1971duality}. Define $\mathbb{Z}_3$ clock degrees of freedom $\mu^z$ and $\mu^x$ on each vertex of the honeycomb lattice, and the following duality tranformations (shown in Fig.~\ref{fig:duality}):
\begin{subequations}
\begin{align}
&\mu_s^z = A_s = \prod_{i\in s} Z_i,  \quad  s \in {\rm sublattice} \ A   \\
&\mu_s^z = A_s^\dagger = \prod_{i \in s} Z_i^\dagger, \quad s \in {\rm sublattice} \ B \\
&\mu_s^x \mu_{\tilde{s}}^{x \dagger} = X_i,  \quad   i \equiv (s\tilde{s}),
\end{align}
\label{eq:duality}
\end{subequations}
where $(s\tilde{s})$ denotes the bond connecting adjacent vertices $s \in$ sublattice $A$ and $\tilde{s} \in$ sublattice $B$.
Notice that a single $\mu^x$ operator is expressed as a string operator in terms of the gauge degrees of freedom:
\begin{equation}
\mu_s^x = \prod_C X_i^\dagger X_j,
\label{eq:string}
\end{equation}
where $C$ denotes an arbitrary path starting from a link emanating
from site $s$ and ending at infinity. The string operator involves
alternating $X$ and $X^\dagger$, and we choose the convention that a
string with endpoint on sublattice $A$ ends with $X$, and a string
with endpoint on sublattice $B$ ends with $X^\dagger$. With this
convention, one can readily check that the dual variables $\mu^x$ and
$\mu^z$ satisfy the correct commutation relations for $\mathbb{Z}_3$
clock variables. In terms of the dual variables,
Hamiltonian~(\ref{eq:dimerization}) maps to
\begin{eqnarray}
H_{\rm Potts} = &&J \sum_i (\mu^z_i + \mu^{z \dagger}_i) - h_s \sum_{\langle i j\rangle \in {\rm vertical}} ( \mu_i^x \mu_{j}^{x\dagger} + \mu_i^{x\dagger} \mu_{j}^{x})    \nonumber    \\
&&- h_w \sum_{\langle i j\rangle \notin {\rm vertical}} ( \mu_i^x \mu_{j}^{x\dagger} + \mu_i^{x\dagger} \mu_{j}^{x}).
\label{eq:Potts}
\end{eqnarray}
Hamiltonian~(\ref{eq:Potts}) describes a $\mathbb{Z}_3$ clock model with ferromagnetic interactions, which is equivalent to a three-state Potts model.

One can further check that Hamiltonians~(\ref{eq:dimerization}) and~(\ref{eq:Potts}) indeed have the same Hilbert space dimension, although naively the quantum double seems to have more degrees of freedom. The Hilbert space dimension of the Potts model~(\ref{eq:Potts}) is $\mathcal{D}_{\rm Potts} = 3^{N_v}$. On the other hand, the quantum double subject to the gauge constraint $G_p=1$ has a Hilbert space dimension of $\mathcal{D}_{\rm gauge} = 3^{N_b-N_p}$. Again using the relations among $N_v$, $N_b$ and $N_p$ on a honeycomb lattice, one finds $\mathcal{D}_{\rm Potts}=\mathcal{D}_{\rm gauge}$.

\subsection{Reduction to quantum XY model}

Based on Hamiltonian~(\ref{eq:Potts}), the system admits a simpler
description in the regime $J \gg h_s, h_w$ that we are interested
in. In this regime, each $\mu^z$ can only take values $\omega$ or
$\overline{\omega}$ in the ground state, which can be modeled as a
two-level system. Define the Pauli spin operator $\sigma^z=+1$ if
$\mu^z=\omega$, and $\sigma^z=-1$ if $\mu^z=\overline{\omega}$. A pair
of nearest-nerighbor spins is flippable under the ferromagnetic term
in Hamiltonian~(\ref{eq:Potts}) only if they are opposite. This leads
to the following low energy effective Hamiltonian:
\begin{eqnarray}
H_{XY} &=& -h_s \sum_{\langle ij \rangle} \sigma_i^x \sigma_j^x (1-\sigma_i^z \sigma_j^z) - h_w \sum_{(ij)} \sigma_i^x \sigma_j^x (1-\sigma_i^z \sigma_j^z)   \nonumber   \\
&=& - h_s \sum_{\langle ij \rangle} ( \sigma_i^x \sigma_j^x + \sigma_i^y \sigma_j^y) - h_w \sum_{(ij)}  ( \sigma_i^x \sigma_j^x + \sigma_i^y \sigma_j^y),  \nonumber   \\
\end{eqnarray}
where we have introduced the short-hand notations $\langle ij \rangle$
and $(ij)$ for a bond that is or is not vertical, respectively. In
particular, at the isotropic point $h_s=h_w$, the system maps to an
isotropic quantum spin-1/2 XY model. The spin-1/2 XY model is known to
have long-range order in the ground state in dimensions greater than
one, and the spectrum is gapless~\cite{PhysRevLett.61.2582,
  PhysRevB.60.6588}. Moreover, in Appendix~\ref{sec:app:bosonization}
we show that the XY phase is stable in the regime $h_s \ll h_w$
starting from weakly coupled chains, using bosonization
techniques.

\begin{figure}[t]
\centering
\includegraphics[width=.3\textwidth]{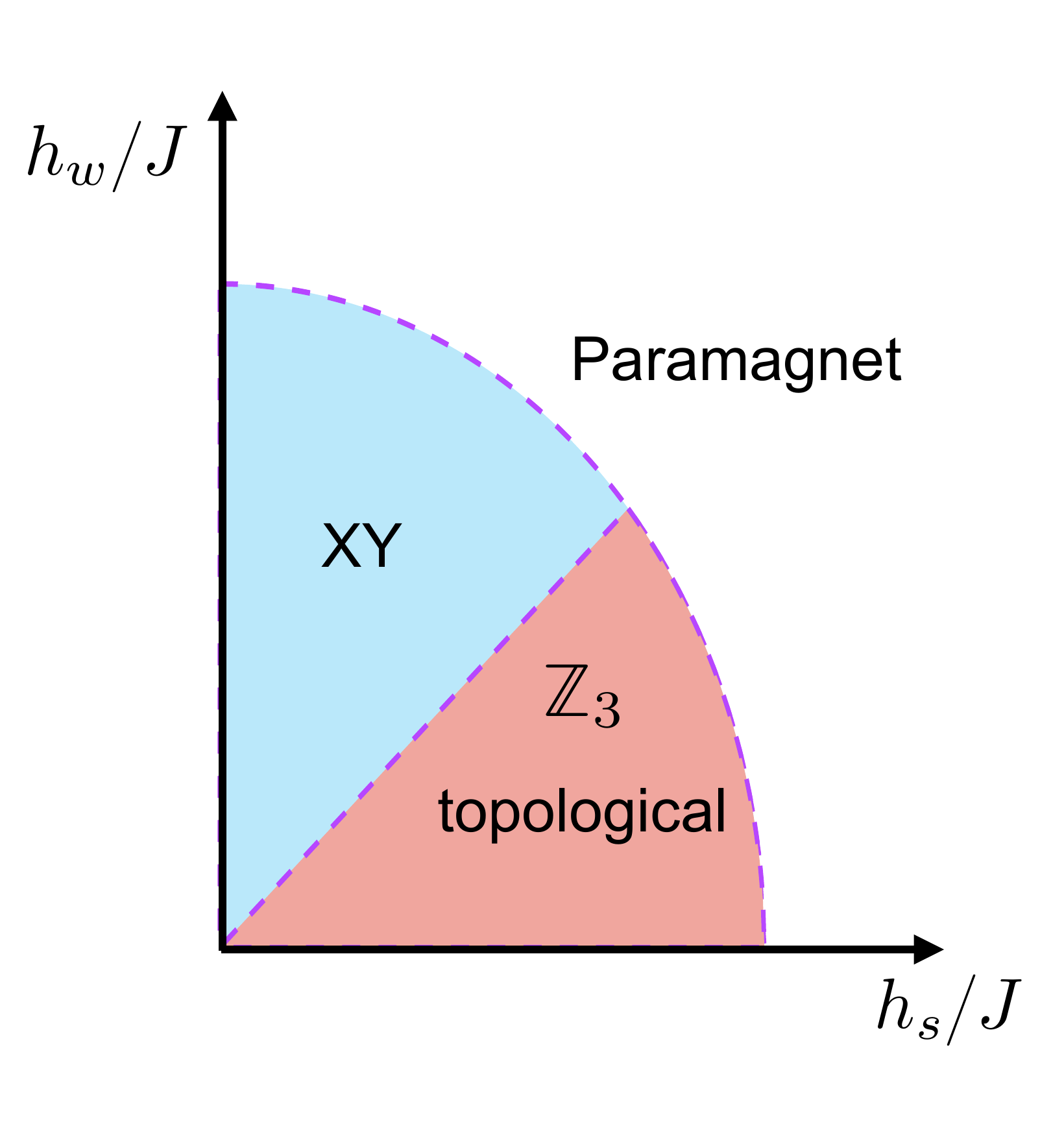}
\caption{A schematic phase diagram of Hamiltonian~(\ref{eq:dimerization}).}
\label{fig:phase_diagram}
\end{figure}

These results underscore the necessity of a strong dimerization with
$h_w<h_s$ to stabilize the ${\mathbb Z}_3$ topological phase, as
discussed above. We conclude with a schematic of the phase diagram of
Hamiltonian~(\ref{eq:dimerization}) in Fig.~\ref{fig:phase_diagram}.

\section{Mean field theory in the bond-operator representation}

\begin{figure}[t]
\centering
\quad \includegraphics[width=0.85\columnwidth]{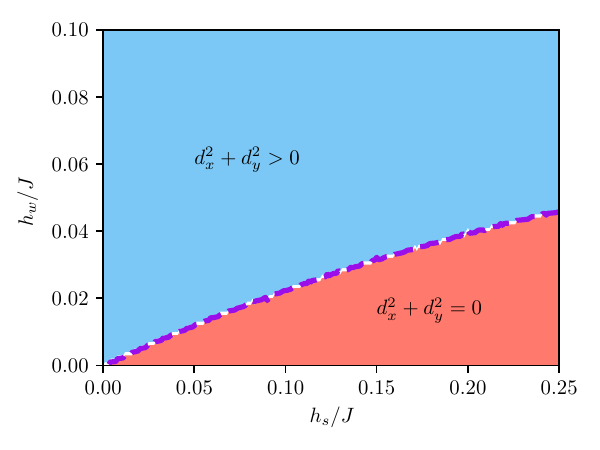}
\begin{center}
(a)
\end{center}
\includegraphics[width=0.9\columnwidth]{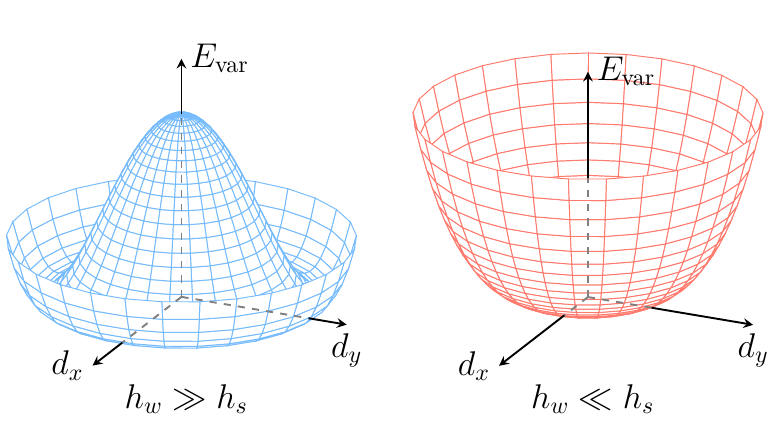}
\begin{center}
(b)
\end{center}
\caption{Numerical results of variational calculations with all nine basis states associated with a bond included in the variational wavefunction. (a) Color plot showing regions in the parameter space where \(\{\ket{\omega\omega}, \ket{\overline{\omega}\overline{\omega}}\}\) components in the ground state wavefunction are zero vs. non-zero. (b) Schematic plots of the variational energy against \(d_x\) and \(d_y\) in the two phases.}
\label{fig:mean_field}
\end{figure}

We now present a mean field theory that is capable of capturing the phase transition between the $\mathbb{Z}_3$ topological phase and the XY-ordered phase. Such a mean field theory is most conveniently formulated in the dual model~(\ref{eq:Potts}). Since $J$ is large in this regime, we can again work with a truncated Hilbert space with two states per site: $|\omega\rangle$ and $|\overline{\omega}\rangle$. From our previous discussions in Sec.~\ref{sec:topological}, deep inside the topological phase, the system essentially forms dimers on the strong bonds, and the vertices within a dimer are strongly entangled. Therefore, we shall formulate our mean field theory using the bond-operator representation~\cite{PhysRevB.41.9323}. The variational wavefunction is then chosen as a tensor product of dimers on the vertical bonds in the bond-operator basis.

The four states in the Hilbert space of a bond can be combined to form singlet and triplet states:
\begin{equation}
\begin{split}
\ket{s} & = \frac{1}{\sqrt{2}} \left(\ket{\omega\overline{\omega}} - \ket{\overline{\omega}\omega}\right), \\
\ket{t_x} & = \frac{-1}{\sqrt{2}} \left(\ket{\omega\omega} - \ket{\overline{\omega}\overline{\omega}}\right), \\
\ket{t_y} & = \frac{i}{\sqrt{2}} \left(\ket{\omega\omega} + \ket{\overline{\omega}\overline{\omega}}\right), \\
\ket{t_z} & = \frac{1}{\sqrt{2}} \left(\ket{\omega\overline{\omega}} + \ket{\overline{\omega}\omega}\right).
\end{split}
\end{equation}
We take the following variational ansatz of the wavefunction
\begin{equation}
\ket{\Psi} = \bigotimes_{\text{dimers}} \left(c_s\ket{s} + c_x \ket{t_x} + c_y \ket{t_y} + c_z \ket{t_z}\right)\ .
\label{eq:variational_ansatz}
\end{equation}
We expect such a variational wavefunction to be a good ansatz for the actual quantum state deep in the topological phase when $h_s \gg h_w$.
The variational energy per unit cell is
\begin{equation}
\begin{split}
&E_{\rm var}= \ev{H_{J\to\infty}}{\Psi}    \\
&= h_s \left(\abs{c_s}^2 - \abs{c_{z}}^2\right) + h_w\left[2\left(\abs{c_s}^2 - \abs{c_z}^2\right)\left(\abs{c_x}^2+\abs{c_y}^2\right) \right.    \\
&\quad \left. + \left(c_x^2+c_y^2\right)\left(\overline{c}_s^2 + \overline{c}_z^2\right) + \left(\overline{c}_x^2+\overline{c}_y^2\right)\left(c_s^2 + c_z^2\right)\right].
\label{eq:j-inf-var-en}
\end{split}
\end{equation}
Notice in the above expression that only terms on the third line
depend on the phases of the variational parameters, while all others
only depend on their norms. Thus, we may choose the phases of the
variational parameters such that the third line is minimized. Let us
define \(c_x^2 + c_y^2 \equiv A = \abs{A} e^{i\phi}\), and \(c_s^2 +
c_z^2 \equiv B = \abs{B} e^{i\theta}\), and rewrite the third line as
\(2 \abs{A} \abs{B} \cos(\phi - \theta)\). This term is minimized when
$\phi-\theta=\pi$, the phases of $c_x$ and $c_y$ are equal, and the
phases of $c_s$ and $c_z$ are equal. One can then use the freedom in
the global U(1) phase of the wavefunction to set both $c_s$ and $c_z$
to be real, which also fixes $c_x$ and $c_y$ to be purely
imaginary. Let us now define real parameters: \(d_s \equiv c_s\),
\(d_z \equiv c_z\), \(d_x \equiv - i c_x\), and \(d_y \equiv - i
c_y\), in terms of which the variational energy becomes
\begin{equation}
E_\text{var} = h_s d_s^2 + (- 4 h_w - h_s) d_z^2 + 4 h_w d_s^2 d_z^2 + 4 h_w d_z^4\,
\end{equation}
where we have used the normalization condition of the wavefunction.
When \(h_s\) and \(h_w\) are both positive, there are always two local minima of the variational energy at
\[
d_s = 0\ , \quad  d_z = \pm \sqrt{\frac{h_s + 4 h_w}{8 h_w}}\ .
\]
However, \(\sqrt{(h_s + 4 h_w) / 8 h_w} > 1\) when \(h_w / h_s < 1/4\), which lies outside the domain \(d_0^2 + d_z^2 \leq 1\).
Therefore, when \(h_w/ h_s < 1/4\), the true minimum is achieved at the boundary, where \(d_s = 0, d_z = \pm 1\).

This gives us the phase transition in the mean field theory. When \(h_w / h_s < 1/4\), the ground state is \(\bigotimes_{\text{dimers}}\ket{t_z} = \bigotimes_{\text{dimers}} (\ket{\omega\overline{\omega}} + \ket{\overline{\omega}\omega})/\sqrt{2}\), consistent with the scenario in the topological phase that we discussed in Sec.~\ref{sec:topological}.
When \(h_w / h_s > 1/4\), the ground state has \(d_s = 0\) and \(d_z\) taking a value less than \(1\), which means that \(d_x^2 + d_y^2\) becomes non-zero. Moreover, the variational energy minimum only depends on $d_x^2+d_y^2$, hence forming a ``Mexican hat''-like profile with O(2) symmetry on top of which an XY-ordered phase can emerge from fluctuations beyond mean field.

While the above simplification in the infinite $J$ limit allows for an elegant analytical treatment for the phase transition, we further perform numerical minimization of the variational energy by including all nine basis states associated with a bond in the variational wavefunction. We use the total weight of the \(\{\ket{\omega \omega}, \ket{\overline{\omega}\overline{\omega}}\}\) components in the ground state wavefunction as an indicator of the phase transition, which is equal to $d_x^2+d_y^2$. The results are shown in Fig.~\ref{fig:mean_field}. Again, we find a critical $h_w$ beyond which a non-zero $d_x^2+d_y^2$ emerges in the ground states, indicating the transition from the topological phase into the XY-ordered phase. Notice that the paramagnetic phase shown in Fig.~\ref{fig:phase_diagram} is absent in the mean field calculations. The paramagnetic phase corresponds to the ferromagnetically ordered phase in the dual model, for which the variational ansatz~(\ref{eq:variational_ansatz}) is no longer a good one. Hence our mean field theory does not capture the transition into the paramagnetic phase.

\section{Summary and outlook}

In this paper we presented a realization of a $\mathbb Z_3$ quantum
double through a Hamiltonian with only physical interations, namely
the Josephson couplings and the capacitances of a superconducting wire
array. The construction hinges on the combinatorial $\mathbb Z_3$
gauge symmetry of the Hamiltonian: both the Josephson and capacitive
terms are invariant under left/right monomial transformations. This
invariance allows the construction of strings of operators that
generate an exact local gauge symmetry.

We discussed in detail the consequences of having an inverted star
potential in the $\mathbb Z_3$ quantum double model, and the
dimerizations that lead to a topologically ordered ground state {\it
  versus} those that stabilize a quantum XY-ordered state. We obtained
the phase diagram of the model as function of parameters $h_s$ and
$h_w$ that microscopically are tied to the capacitances and the
Josephson energy scale $J$. We show that another consequence of the
inverted star potential is that the vison gap can be larger than that
in the uninverted case, as it occurs to lower order in perturbation
theory and as a function of a larger dimensionless ratio ($h_w/h_s$,
instead of $h_{s,w}/J$).

Our work opens fronts to tackle the problem of realizing quantum
double models with realistic interactions that span beyond the
specific construction for the group $\mathbb Z_3$ in superconducting
arrays. As a simple example, once one obtains the Hadamard matrices
using the complex numbers $\omega,\overline{\omega}$ (that originate
from fluxes in the superconducting realization), one can easily
construct spin-1/2 systems with one- and two-body interactions with
the necessary combinatorial symmetry to realize the same quantum
double. It then remains to be investigated whether the model supports
a gapped phase with \(\mathbb{Z}_3\) topological order.  That this
spin representation is possible follows from replacing these complex
numbers $1, \omega,\overline{\omega}$ by their $3\times 3$ permutation
representations:
\begin{align}
  1\to
  \begin{pmatrix}
   1&0&0\\0&1&0\\0&0&1
  \end{pmatrix}
  ,
  \quad
  \omega \to
  \begin{pmatrix}
   0&1&0\\0&0&1\\1&0&0
  \end{pmatrix}
  ,
  \quad
  \overline{\omega}\to
  \begin{pmatrix}
   0&0&1\\1&0&0\\0&1&0
  \end{pmatrix}
  \;.
\end{align}
The \(W\) matrix then becomes a \(9\times 9\) matrix,
  invariant under pairs of left/right monomial transformations as in
  \eqref{eq:auto}, which are now represented by \(9\times 9\)
  permutation matrices.  This matrix of interactions corresponds to
$ZZ$ spin interactions between 9 matter spins at the sites of the
honeycomb lattice, with 3 gauge spins at each of the links emanating
from each site.  While this may appear an unlikely model to encounter
in nature, we stress that these kinds of couplings are the same as
those used to embed the $\mathbb Z_2$ model in the D-Wave DW-2000Q
quantum device~\cite{Zhou2020}. Embedding the $\mathbb Z_3$ model in
such devices is not unrealistic, specially if one explores newer
architectures with larger qubit connectivities, such as those in the
D-Wave Advantage device.

On yet a different level, the successful construction of the $\mathbb
Z_3$ quantum double on top of combinatorial gauge symmetry is not an
end on itself, but simply points to the promise that other quantum
doubles -- Abelian and, more interestingly, non-Abelian -- could be
constructed. The search for realistic models with at most two-body
interactions acquires a systematic path: one must first find coupling
matrices with elements in a given group $G$ that are invariant under
multiplication on the left/right by monomial matrices with elements in
$G$. If the condition is further satisfied by right matrices that are
diagonal, with only two of the elements along the diagonal not equal to
1, loops can be constructed defining a local gauge symmetry. Once this
abstract step of constructing such coupling matrices succeeds, one can
find a monomial representation of the group elements and consequently
translate the abstract model to a spin Hamiltonian with at most
two-body interactions. The pursuit of this generic pathway to
constructing quantum doubles for different groups is a possibility
that this paper raises.

\section*{Acknowledgments}

We thank Andrew J. Kerman for a discussion on superconducting arrays
on the honeycomb lattice that stimulated this work. We thank Garry
Goldstein and Andrei Ruckenstein for constructive criticism and useful
discussions. In particular, we thank Garry Goldstein for pointing out
to us a correction to the calculation in
Eq.~\ref{eq:supset}. Z.-C. Y. would like to thank Jyong-Hao Chen for
useful exchanges on bosonization. Z.-C. Y. acknowledges funding by the
DoE ASCR Accelerated Research in Quantum Computing program (award
No. DE-SC0020312), U.S. Department of Energy Award No. DE-SC0019449,
DoE ASCR Quantum Testbed Pathfinder program (award No. DE-SC0019040),
NSF PFCQC program, AFOSR, ARO MURI, AFOSR MURI, and NSF PFC at JQI.
Z.-C. Y. is also supported by MURI ONR N00014-20-1-2325, MURI AFOSR,
FA9550-19-1-0399, and Simons Foundation. This work was supported in
part by DOE Grant No. DE-FG02-06ER46316 (H. Y. and C. C.) and by NSF
Grant DMR-1906325 (C. C.).

\appendix

\section{Minimum of the Josephson potential}
\label{sec:app:phases}

We seek the minimum of the Josephson potential
\begin{equation}
  -E_J \sum_s \left[ \sum_{i,a \in s} W_{ai} \ e^{i(\theta_i-\phi_a)} + {\rm h.c.}\right]
  \;.
\end{equation}
Let us denote $z_i=e^{i\theta_i}$ and $v_a=e^{i\phi_a}$. The potential minima, subject to the constraint $|z_i|^2=|v_a|^2=1$, can be found by minimizing the function
\begin{eqnarray}
F = &-& E_J \sum_{ia} \left( z_i \ W_{ai}\ v_a^* + v_a \ W_{ia}^* \ z_i^* \right)   \nonumber   \\
&-& \sum_i \lambda_i \left(|z_i|^2-1\right) - \sum_a \gamma_a \left(|v_a|^2-1\right),
\end{eqnarray}
where $\lambda_i$ and $\gamma_a$ are Lagrange multipliers. Taking the derivative with respect to $v_a^*$ yields
\begin{equation}
\frac{\partial{F}}{\partial{v_a^*}} = -E_J \sum_i z_i W_{ai} - \gamma_a v_a = 0.
\label{eq:F}
\end{equation}
Using the fact that $v_a$ is a pure phase, and $\gamma_a$ is real, we obtain
\begin{equation}
|\gamma_a| = E_J \left|\sum_i z_i W_{ai}\right|,  \quad  {\rm and}  \quad   \gamma_a = -E_J \sum_i z_i \ W_{ai} \ v_a^*.
\end{equation}
The minimal energy can be written as
\begin{eqnarray}
E_{\rm min} &=& -E_J \sum_{ia} \left( z_i \ W_{ai}\ v_a^* + v_a \ W_{ia}^* \ z_i^* \right) = 2\sum_a \gamma_a  \nonumber  \\
&\geq& -2 \sum_a |\gamma_a| = -2E_J \sum_a \left| \sum_i z_i W_{ai} \right|.
\label{eq:minima}
\end{eqnarray}
Eq.~(\ref{eq:minima}) implies that the Josephson energy minima are
given by the gauge wire phase configurations $(\theta_1, \theta_2,
\theta_3)$ such that the potential
\begin{equation}
  -2E_J \sum_a \left|\sum_i W_{ai}
  e^{i\theta_i} \right|
\end{equation}
is minimized. Notice from Eq.~(\ref{eq:F}) that the matter wire phases
$\phi_a$ are completely tethered to $\theta_i$. Using the fact
  that $\gamma_a<0$ at the minima, one can solve for $\phi_a$ for a
  given set of $\theta_i$ via
\begin{equation}
e^{i\phi_a} = \frac{\sum_i W_{ai} e^{i\theta_i}}{\left|\sum_i W_{ai} e^{i\theta_i}\right|}.
\label{eq:phi-appendix}
\end{equation}

\section{Estimate of the tunneling amplitude from Euclidean action of the instanton}
\label{app:instanton}

We estimate the amplitude for tunneling between adjacent minima as shown in Fig.~\ref{fig:potential}. As we discussed in Sec.~\ref{sec:minima}, such processes correspond to shifting the superconducting phases of one gauge wire by $\pm \frac{2\pi}{3}$ while keeping the other two unchanged, thus it gives an estimate for the transverse field strength in the effective $\mathbb{Z}_3$ Hamiltonian~(\ref{eq:clock}).

The Euclidean action for a single waffle is written as:
\begin{eqnarray}
S_E \left[ \theta, \phi; \dot{\theta}, \dot{\phi} \right] &=& \int_{\tau_i}^{\tau_f} \mathcal{L}_E(\theta, \phi; \dot{\theta},  \dot{\phi}) \ d\tau  \nonumber  \\
&=&  \int_{\tau_i}^{\tau_f} \left[ K(\dot{\theta}, \dot{\phi}) + V(\theta, \phi) \right] \ d\tau,
\end{eqnarray}
where the potential energy
\begin{equation}
V(\theta, \phi) = -E_J \sum_{i,a} \left( W_{ai} \ e^{i(\theta_i-\phi_a)} + {\rm h.c.} \right),
\end{equation}
and the kinetic energy
\begin{eqnarray}
K(\theta, \phi) = &&\frac{1}{2} C_g \sum_{i=1}^3 \dot{\theta}_i^2 + \frac{1}{2} C_m \sum_{a=1}^3 \dot{\phi}_a^2 + \frac{1}{2} C_J \sum_{i,a} \left(\dot{\theta}_i - \dot{\phi}_a \right)^2  \nonumber   \\
&&+ \ \frac{1}{2} C_p \left[ \left(\dot{\theta}_1 - \dot{\theta}_2 \right)^2 + \left( \dot{\theta}_2 - \dot{\theta}_3\right)^2 \right]  \nonumber  \\
&&+ \ \frac{1}{2} C_p \left[ \left(\dot{\phi}_1 - \dot{\phi}_2 \right)^2 + \left( \dot{\phi}_2 - \dot{\phi}_3\right)^2 \right].
\end{eqnarray}
As we mentioned in the main text, a small value of $C_p$ breaks the permutation symmetry among the three matter
wires; nevertheless, if Hamiltonian~(\ref{eq:model}) supports a gapped phase with $\mathbb{Z}_3$ topological order, it will remain stable in the presence of a small combinatorial symmetry breaking perturbation so long as the gap stays open.

Due to the combinatorial symmetry, it suffices to consider one particular tunneling process, e.g. the horizontal arrow depicted in Fig.~\ref{fig:potential}(a) where $\theta_2$ changes from $-\frac{2\pi}{3}$ to 0, and $\theta_1=0$, $\theta_3=-\frac{2\pi}{3}$. Throught the tunneling process, all three $\phi$'s will change. However, their trajectories are completely fixed by that of the varying $\theta$, following from Eq.~(\ref{eq:phi}). Thus, one may write
\begin{equation}
\dot{\phi}_a = \frac{d \phi_a}{d \theta} \ \dot{\theta},
\end{equation}
where we have suppressed the gauge wire subscript in $\theta$. The Lagrangian can be simplified as
\begin{widetext}
\begin{eqnarray}
\mathcal{L}_E(\theta, \dot{\theta}) &=& \frac{1}{2} C_g \dot{\theta}^2 + \frac{1}{2} C_m \sum_{a=1}^3 \left( \frac{d \phi_a}{d \theta}\right)^2 \dot{\theta}^2 + \frac{1}{2} C_J \sum_{a=1}^3 \left( \frac{d \phi_a}{d \theta}  \right)^2 \dot{\theta}^2 \times 2 + \frac{1}{2} C_J \sum_{a=1}^3 \left(1 - \frac{d \phi_a}{d \theta} \right)^2 \dot{\theta}^2 + V_{\rm min}( \theta)  \nonumber   \\
&=& \frac{1}{2} \left[ C_g + \sum_{a=1}^3 \left( \frac{d \phi_a}{d \theta}\right)^2 (C_m+2C_J) + \sum_{a=1}^3 \left(1 - \frac{d \phi_a}{d \theta} \right)^2 C_J \right]  \dot{\theta}^2  + V_{\rm min}(\theta)    \nonumber   \\
&\equiv& \frac{1}{2} C_{\rm eff} \dot{\theta}^2 + V_{\rm min}(\theta),
\end{eqnarray}
\end{widetext}
where we have defined an effective capacitance $C_{\rm eff}$, and $V_{\rm min}$ is given by Eq.~(\ref{eq:minima}), which is the profile plotted in Fig.~\ref{fig:potential}(a). A particle initially at one minimum of $V_{\rm min}$ has energy $E_{\rm min} =  -6 E_J$. From energy conservation (in Euclidean space), one obtains:
\begin{equation}
\dot{\theta} = \sqrt{\frac{2[V_{\rm min}(\theta) + 6 E_J]}{C_{\rm eff}}}.
\end{equation}
Hence, the Euclidean action corresponding to this classical trajectory is given by
\begin{equation}
S_E = \int_{\theta_i=-\frac{2\pi}{3}} ^{\theta_f = 0}  \sqrt{2 C_{\rm eff} [V_{\rm}(\theta) + 6E_J]} \ d \theta,
\end{equation}
and the tunneling amplitude is $\sim e^{-S_E}$.

In principle, the effective potential $C_{\rm eff}$ is not a constant along the trajectory, due to the $\theta$-dependence in $d \phi_a/d \theta$. Nevertheless, a straightforward calculation of Eq.~(\ref{eq:phi}) yields the following simple relations between $\phi_a$ and $\theta$:
\begin{subequations}
\begin{align}
\phi_1 &= \frac{1}{2} \theta - \frac{\pi}{6},   \\
\phi_2 &= \frac{1}{2} \theta + \frac{\pi}{2},  \\
{\rm tan} \phi_3 & = \frac{{\rm sin}\left( \theta-\frac{2\pi}{3} \right)}{2+{\rm cos} \left( \theta-\frac{2\pi}{3} \right)}.
\end{align}
\end{subequations}
We find that both $d \phi_1/d \theta$ and $d \phi_2/ d \theta$ are in fact constant. Therefore as an approximation, we may take $C_{\rm eff}$ to be a constant along the trajectory.
Evaluating the action numerically yields the tunneling amplitude $\sim e^{-0.88 \sqrt{C_{\rm eff} E_J}}$.

\begin{figure}[t]
\centering
\includegraphics[width=.35\textwidth]{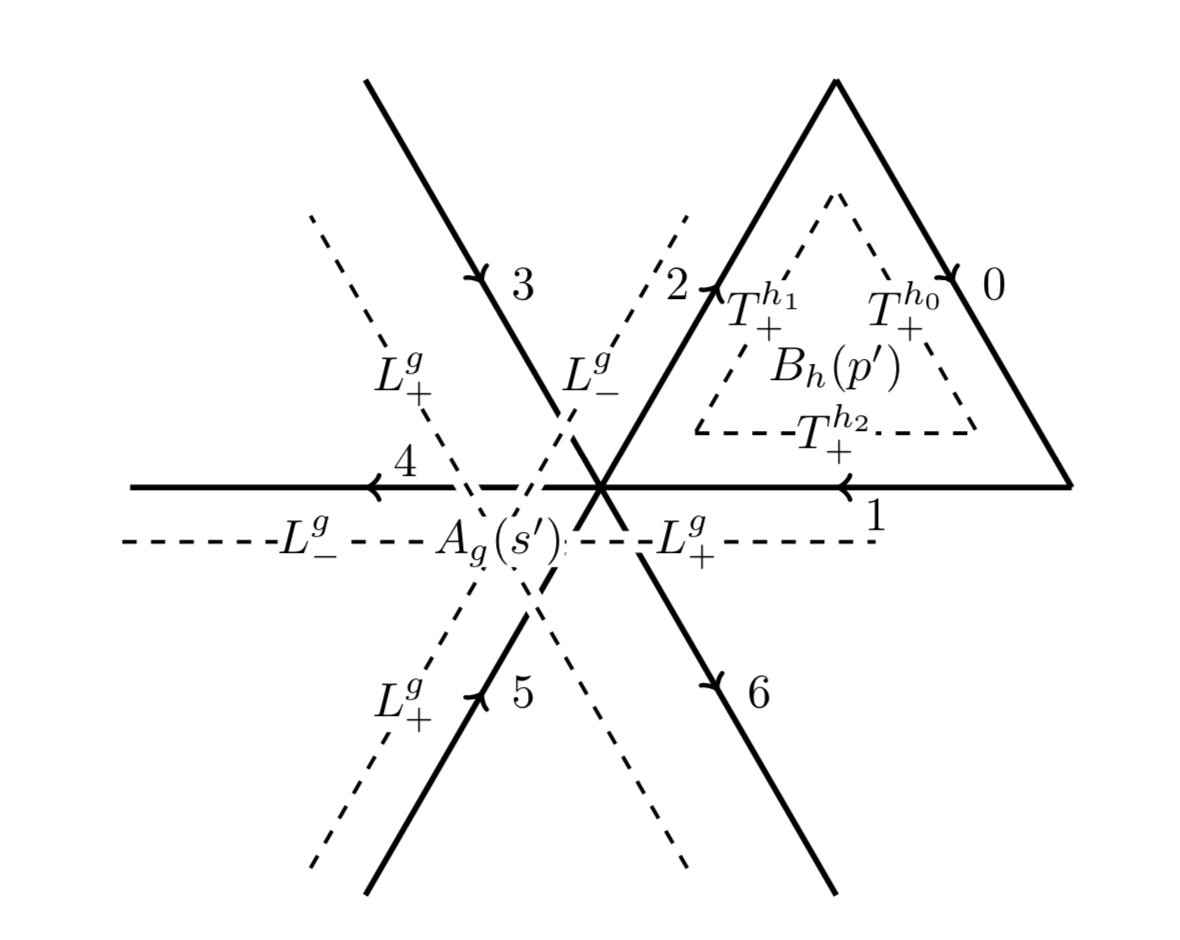}
\caption{Formulation of the quantum double model $\mathfrak{D}(\mathbb{Z}_3)$ on the dual triangular lattice. The arrows indicate the orientation rules of the lattice.}
\label{fig:quantum_double}
\end{figure}

\section{Equivalence between Hamiltonian~(\ref{eq:clock}) and quantum double model}
\label{app:equivalence}

We show that Hamiltonian~(\ref{eq:clock}) with the gauge constraint is equivalent to Kitaev's quantum double model $\mathfrak{D}(\mathbb{Z}_3)$ in the non-zero flux sector. Since our gauge transformation $G_p$'s are defined on the plaquettes, the corresponding quantum double model is most conveniently formulated on the dual triangular lattice, as depicted in Fig.~\ref{fig:quantum_double}.

Let us define an orthonormal basis on each link of the triangular lattice: $\{ | 1\rangle, |\omega \rangle, |\overline{\omega}\rangle \}$. The construction of $\mathfrak{D}(\mathbb{Z}_3)$ starts from the following group-element-indexed linear operators acting on the above Hilbert space~\cite{kitaev2003fault}
\begin{equation}
L^g_+ |z\rangle = |gz\rangle \quad T^h_+ |z\rangle = \delta_{h,z} |z\rangle,
\end{equation}
where $g, h, z \in \mathbb{Z}_3$. And similarly, one can define $L^g_-$ and $T^h_-$, which, for abelian groups, are simply $L^g_- = L^{g^{-1}}_+$ and $T^h_- = T^{h^{-1}}_+$. In terms of the clock operators, $L^g_{\pm}$ and $T^h_{\pm}$ have the explicit form
\begin{widetext}
\begin{subequations}
\begin{align}
L^I_+ = I  \quad  &L^{\omega}_+ = X \quad  L^{\overline{\omega}}_+ = X^2,    \\
T^I_+ = \frac{1}{3} (I + Z + Z^2)    \quad   T^{\omega}_+ = \frac{1}{3}( I + &\overline{\omega} Z + \omega Z^2)  \quad  T^{\overline{\omega}}_+ = \frac{1}{3} (I + \omega Z + \overline{\omega} Z^2).
\end{align}
\end{subequations}
\end{widetext}
One can further check that the above operators $L^g_{\pm}$ and $T^h_{\pm}$ satisfy the commutation relation
\begin{equation}
L^g_+ T^h_+ = T^{gh}_+ L^g_+,
\end{equation}
from which all other commutation relations involving $L^g_{\pm}$ and $T^g_{\pm}$ follow. We further choose an orientation rule on the triangular lattice as depicted in Fig.~\ref{fig:quantum_double}, such that the arrows go clockwise (counterclockwise) around every upward (downward) pointing triangle. For each vertex $s$ and the bonds emanating from $s$, we take $L^g_-$ if the arrow is pointing towards $s$, and $L^g_+$ otherwise. For each plaquette $p$ and the bonds surrounding $p$, we take $T^h_-$ if $p$ is to the left of the bond following the arrow, and $T^h_+$ otherwise. Using the above rules, one can construct the generators of $\mathfrak{D}(\mathbb{Z}_3)$ as follows
\begin{widetext}
\begin{subequations}
\begin{align}
A_I(s) = I  \quad  &A_{\omega}(s) = X_1 X_2^\dagger X_3 X_4^\dagger X_5 X_6^\dagger   \quad     A_{\overline{\omega}} = X_1^\dagger X_2 X_3^\dagger X_4 X_5^\dagger X_6,  \\
B_I(p) = \frac{1}{3} (I + Z_0 Z_1 Z_2 + Z_0^\dagger Z_1^\dagger Z_2^\dagger)   \quad   B_{\omega}(p) &= \frac{1}{3}(I + \overline{\omega} Z_0 Z_1 Z_2 + \omega Z_0^\dagger Z_1^\dagger Z_2^\dagger)   \quad  B_{\overline{\omega}}(p) = \frac{1}{3}( I + \omega Z_0 Z_1 Z_2 + \overline{\omega} Z_0^\dagger Z_1^\dagger Z_2^\dagger),
\end{align}
\end{subequations}
\end{widetext}
where the labels are shown in Fig.~\ref{fig:quantum_double}. In terms of the above generators, one can write down the star term, which has the form of a projector~\cite{kitaev2003fault}:
\begin{equation}
A(s) = \frac{1}{3} \left[ I + A_{\omega}(s) + A_{\overline{\omega}}(s)\right],
\end{equation}
and the plaquette term
\begin{equation}
B(p) = B_{\omega}(p) + B_{\overline{\omega}}(p).
\end{equation}
Notice that in the usual quantum double model, the plaquette term enforces a zero flux: $B(p)=B_I(p)$. Here $B(p)$ instead favors sectors with flux $\omega$ or $\overline{\omega}$, which corresponds to the inverted potential in Hamiltonian~(\ref{eq:clock}). Finally, we can write down the Hamiltonian for the quantum double model $\mathfrak{D}(\mathbb{Z}_3)$:
\begin{equation}
H = \sum_s \left[ 1- A(s) \right] + \sum_p \left[ 1-B(p) \right].
\end{equation}
Going back from the dual triangular lattice to the honeycomb lattice, this is precisely Hamiltonian~(\ref{eq:clock}) in the absence of a transverse field and with the gauge constraint imposed.

\begin{figure}[t]
\centering
\includegraphics[width=.35\textwidth]{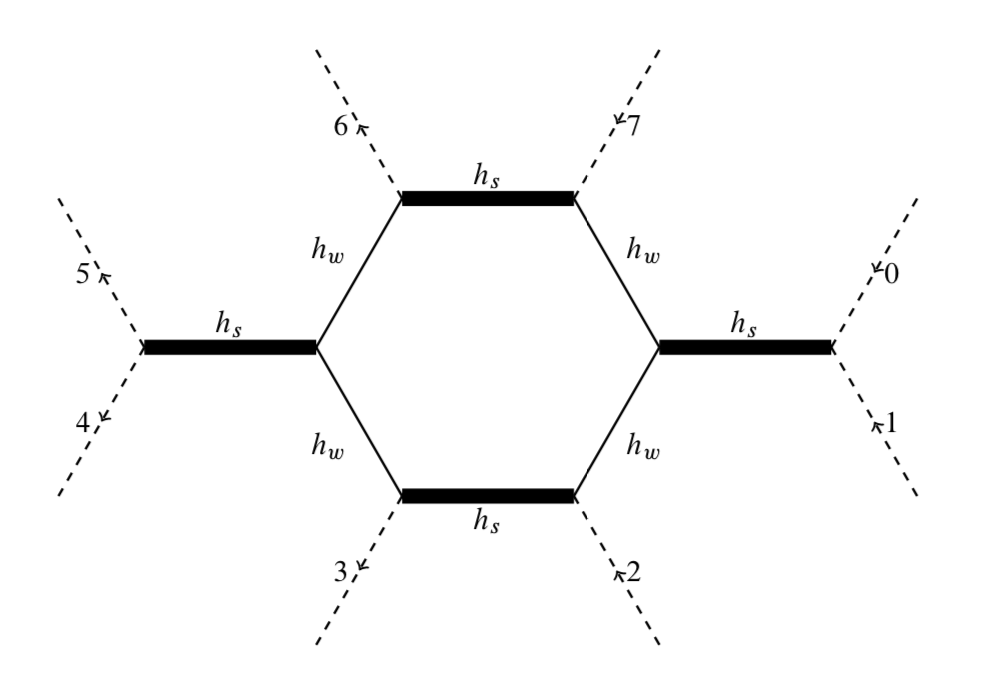}
\includegraphics[width=0.35\textwidth]{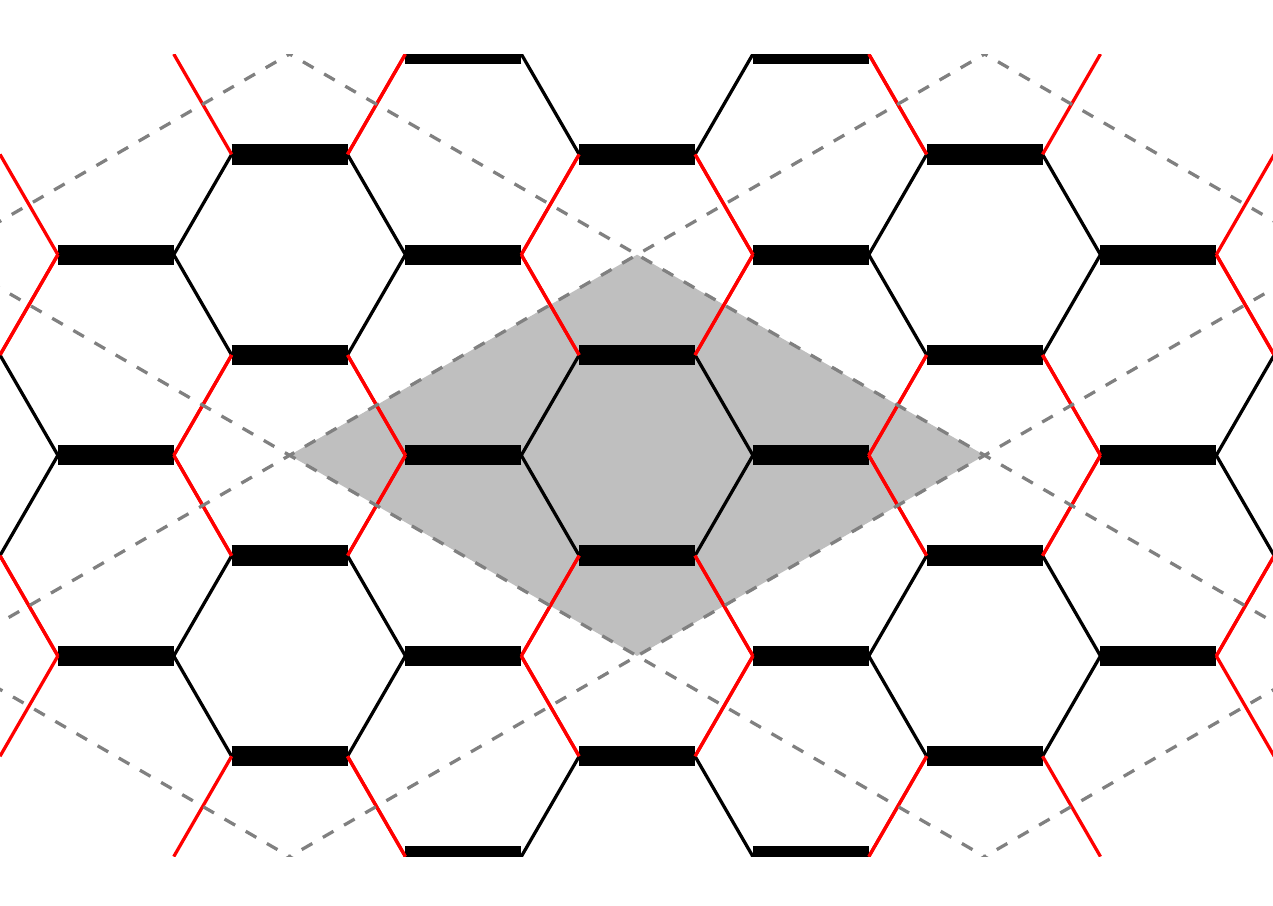}
\caption{The ``spider'' like geometry considered in exact diagonalization, and its tiling of the entire lattice.}
\label{fig:spider}
\end{figure}

\section{Numerical results of Hamiltonian~(\ref{eq:dimerization}) on a ``spider'' like geometry}
\label{app:spider}

We show exact diagonalization results of Hamiltonian~(\ref{eq:dimerization}) on an elementary ``spider'' like geometry depicted in Fig.~\ref{fig:spider}. Since one can tile the entire two dimensional honeycomb lattice using the spider as an elementary building block, the spider can be thought of as a minimal lattice on which one can test our model numerically.

As shown in Fig.~\ref{fig:spider}, we fix the 8 external leg configurations $\{ Z_0, Z_1, \ldots, Z_7\}$, and diagonalize the spectrum of the 8 internal clock degrees of freedom under Hamiltonian~(\ref{eq:dimerization}). We can interpret this particular setup as a single plaquette embedded in the lattice environment, whose configurations are fixed one at a time. There are in total $3^8$ possible external leg configurations that one can fix to. From energetic considerations in the regime $h_w \ll h_s \ll J$ as we discussed in Sec.~\ref{sec:topological}, the ground state forms dimers on the strong bonds. Therefore, we expect that the ground state energy of the spider is minimized when
\begin{equation}
\prod_{s \in A} A_s \prod_{s \in B} A^\dagger_s = 1.
\label{eq:spider_constraint}
\end{equation}
Notice that for the entire system on a torus, the above equation is an identity that imposes a constraint on the spectrum; here it arises from energetics instead. Applying Eq.~(\ref{eq:spider_constraint}) on a spider, we obtain
\begin{equation}
Z_3 Z_4 Z_5 Z_6 Z_0^\dagger Z_1^\dagger Z_2^\dagger Z_7^\dagger = 1.
\label{eq:spider_constraint2}
\end{equation}
Out of the $3^8$ external leg configurations, Eq.~(\ref{eq:spider_constraint2}) yields $3^7$ configurations such that the ground state energy is minimized. We have tested that fixing the external legs to be any of the $3^7$ configurations satisfying Eq.~(\ref{eq:spider_constraint2}) yields the same ground state and first excited state energies, which is a direct consequence of the gauge symmetry.

\begin{figure}[t]
\centering
\includegraphics[width=0.9\columnwidth]{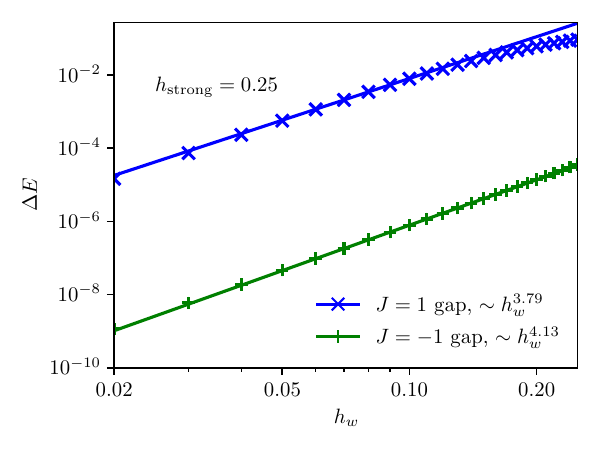}
\caption{Energy gap between the ground state and first excited state of Hamiltonian~(\ref{eq:dimerization}) on a spider as a function of $h_w$, for fixed $h_s=0.25$ and $J = \pm 1$.}
\label{fig:spider_gap}
\end{figure}

In Fig.~\ref{fig:spider_gap}, we plot the energy gap between the ground state and the first excited state as a function of $h_w$, for fixed $J = 1$ and $J = -1$, $h_s$ and external leg configuration satisfying Eq.~(\ref{eq:spider_constraint2}). This can be viewed as the vison gap obtained numerically from a spider building block. In our model where \(J>0\), we find that the fitted gap scales as $\Delta E \sim h_w^{3.82}$ for small $h_w$, which is consistent with our perturbative calculations in Sec.~\ref{sec:topological}.
As a comparison, we also plot in Fig.~\ref{fig:spider_gap} the energy gap for Hamiltonian~(\ref{eq:dimerization}) with $J<0$, which corresponds to the conventional quantum double model with zero flux. In this case, we find a much smaller vison gap than in Fig.~\ref{fig:spider_gap}(a) with an inverted potential. The plaquette term for $J<0$ is generated at sixth order in perturbation theory, which leads to a small vison gap.

\section{Weakly coupled chain limit: bosonization}
\label{sec:app:bosonization}

Another interesting regime that can be understood is when $h_s \ll h_w$. In the limit when $h_s=0$, the system becomes a set of decoupled chains extending along the horizontal direction, as can be seen from Fig.~\ref{fig:dimerization}(a). Since each chain is described by an XY model, which is equivalent to free fermions in one dimension, the system is apparently gapless in this limit. A weak $h_s$ introduces inter-chain couplings along the vertical direction. We shall now study the effect of this inter-chain coupling using abelian bosonization.

\begin{figure}[b]
\centering
\includegraphics[width=.45\textwidth]{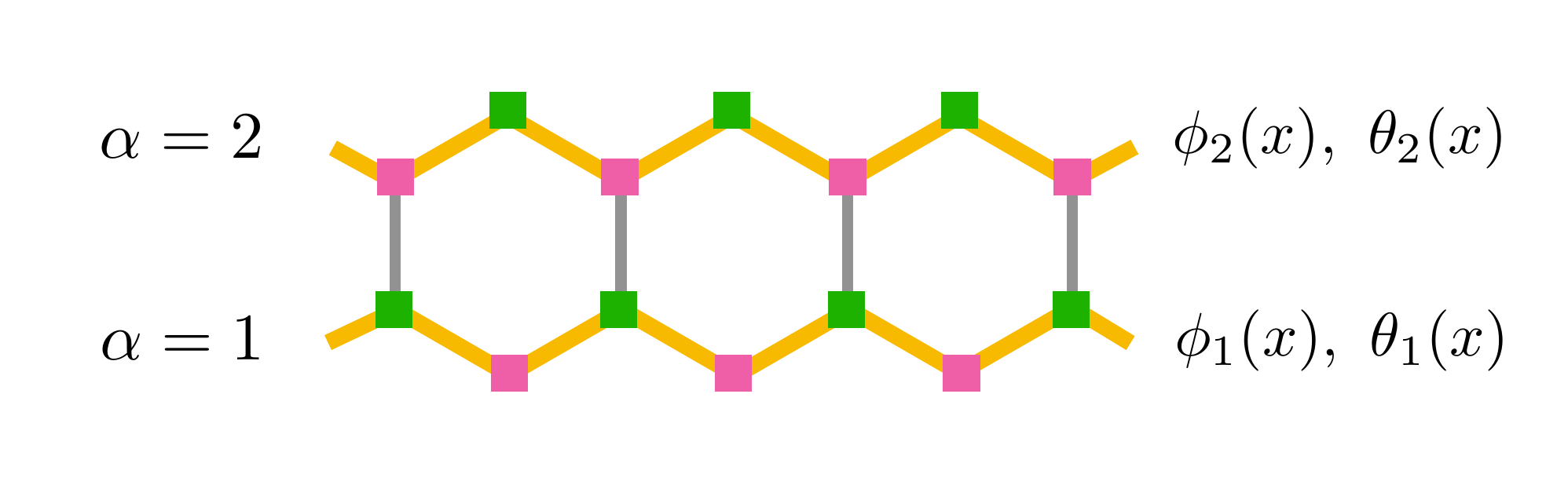}
\caption{A two-leg ladder considered in bosonization. Each leg is described in terms of a bosonic field $\phi_\alpha$ and its dual $\theta_\alpha$ (not to be confused with the superconducting phases).}
\label{fig:bosonization}
\end{figure}

Consider a two-leg ladder shown in Fig.~\ref{fig:bosonization}, which is described by the following Hamiltonian:
\begin{equation}
H = H_1 + H_2 + H_{\perp},
\end{equation}
where $H_\alpha$ describes the decoupled chain for $\alpha = 1, 2$. In terms of bosonic fields, the bosonized decoupled chain Hamiltonian can be written as:
\begin{equation}
H_{\alpha} = \frac{1}{2} \int dx \left[ (\partial_x \phi_\alpha)^2 + (\partial_x \theta_\alpha)^2\right],
\end{equation}
where $\theta_\alpha$ is the dual variable of $\phi_\alpha$ (not to be confused with the superconducting phases). We have ignored the Luttinger parameter $K=1$, as well as a prefactor of $v_F \propto h_w$. To derive the bosonized form of the inter-chain coupling $H_{\perp}$, we need the bosonized form of the spin operators. First, recall the Jordan-Wigner transformation:
\begin{subequations}
\begin{align}
\sigma_j^\dagger &= e^{i \pi \sum_{i<j} \psi^\dagger_i \psi_i} \psi^\dagger_j  \\
\sigma_j^- &= \psi_j e^{-i \pi \sum_{i<j} \psi^\dagger_i \psi_i}.
\end{align}
\end{subequations}
In the continuum limit, the fermion operator expanded near $\pm k_F$ can be written as:
\begin{equation}
\psi(x) \approx e^{i k_F x} \psi_R(x) + e^{-i k_F x} \psi_L(x),
\end{equation}
where $\psi_{R/L}(x)$ describes right and left movers. Finally, we need the following bosonization dictionary~\cite{fradkin2013field, gogolin2004bosonization}:
\begin{subequations}
\begin{align}
\psi_R(x) &= \frac{1}{\sqrt{2\pi a}} e^{-i\sqrt{\pi}(\phi+\theta)},  \\
 \psi_L(x) &= \frac{1}{\sqrt{2\pi a}} e^{-i\sqrt{\pi}(\theta-\phi)},  \\
 \rho(x) &= \rho_0 + \frac{1}{\sqrt{\pi}} \partial_x \phi(x),
\end{align}
\end{subequations}
where we have suppressed the chain index $\alpha$ for now. Using the above expressions, we can now derive~\cite{giamarchi2003quantum, PhysRevB.67.064419}:
\begin{eqnarray}
&\sigma^\dagger(x)&\rightarrow  e^{i \pi \int dx \rho(x)} \psi^\dagger(x)   \nonumber  \\
&=& \frac{e^{i k_F x + i \sqrt{\pi} \phi(x)}}{\sqrt{2\pi a}} \left[ e^{-ik_F x} e^{i \sqrt{\pi} (\phi+\theta)} + e^{i k_F x} e^{i\sqrt{\pi}(\theta-\phi)}\right]   \nonumber   \\
&=& \frac{e^{i \sqrt{\pi} \theta}}{\sqrt{2\pi a}} \left[ (-1)^x + {\rm cos} (2\sqrt{\pi} \phi) \right],
\end{eqnarray}
and similarly for $\sigma^-(x)$. The inter-chain coupling:
\begin{equation}
H_{\perp} = -h_s \sum_i \sigma^\dagger_{i,1} \sigma^-_{i,2} + {\rm h. c.}
\end{equation}
can now be readily bosonized. Introducing the following new variables corresponding to the symmetric and antisymmetric sectors:
\begin{equation}
\phi_{\pm} =\frac{1}{\sqrt{2}}(\phi_1\pm \phi_2),  \quad  \theta_{\pm} = \frac{1}{\sqrt{2}}(\theta_1 \pm \theta_2),
\end{equation}
the full Hamiltonian can be written as:
\begin{equation}
H = H_+ + H_- + H_{\rm couple},
\end{equation}
where
\begin{subequations}
\begin{align}
H_+ &= \frac{1}{2} \left[(\partial_x \phi_+)^2 + (\partial_x \theta_+)^2 \right],    \\
H_- &= \frac{1}{2} \left[(\partial_x \phi_-)^2 + (\partial_x \theta_-)^2 \right]  + \frac{1}{\pi a} {\rm cos} (\sqrt{2\pi} \theta_-),    \\
H_{\rm couple} &= \frac{1}{2\pi a} {\rm cos} (\sqrt{2\pi} \theta_-) \ {\rm cos} (2\sqrt{2\pi} \phi_+)   \nonumber  \\
&\ +\frac{1}{2\pi a} {\rm cos} (\sqrt{2\pi} \theta_-) \ {\rm cos} (2\sqrt{2\pi} \phi_-).
\end{align}
\end{subequations}

In the above expressions, we only keep the slowly varying, non-staggered contributions. We find that a ${\rm cos}(\sqrt{2\pi}\theta_-)$ term is generated in the antisymmetric sector by the inter-chain couplings. This term has a scaling dimension of $\Delta = (\sqrt{2\pi})^2/4\pi = 1/2$, which is relevant. Thus, the antisymmetric sector $H_-$ becomes gapped. To determine the fate of the $H_+$ sector, we may replace ${\rm cos}(\sqrt{2\pi}\theta_-)$ by its expectation value:
\begin{equation}
\lambda \equiv \frac{1}{2\pi a} {\rm cos}(\sqrt{2\pi}\theta_-).
\end{equation}
Then the ${\rm cos}(2\sqrt{2\pi}\phi_+)$ term has a scaling dimension of $\Delta = (2\sqrt{2\pi})^2/4\pi = 2$, which is marginal. Therefore, we find that the system should remain gapless for a non-zero but weak $h_s$.

\section{A $\mathbb{Z}_3$ quantum double with an uninverted star term}
\label{sec:app:uninverted}

\begin{figure*}[t]
\centering
\includegraphics[width=.8\textwidth]{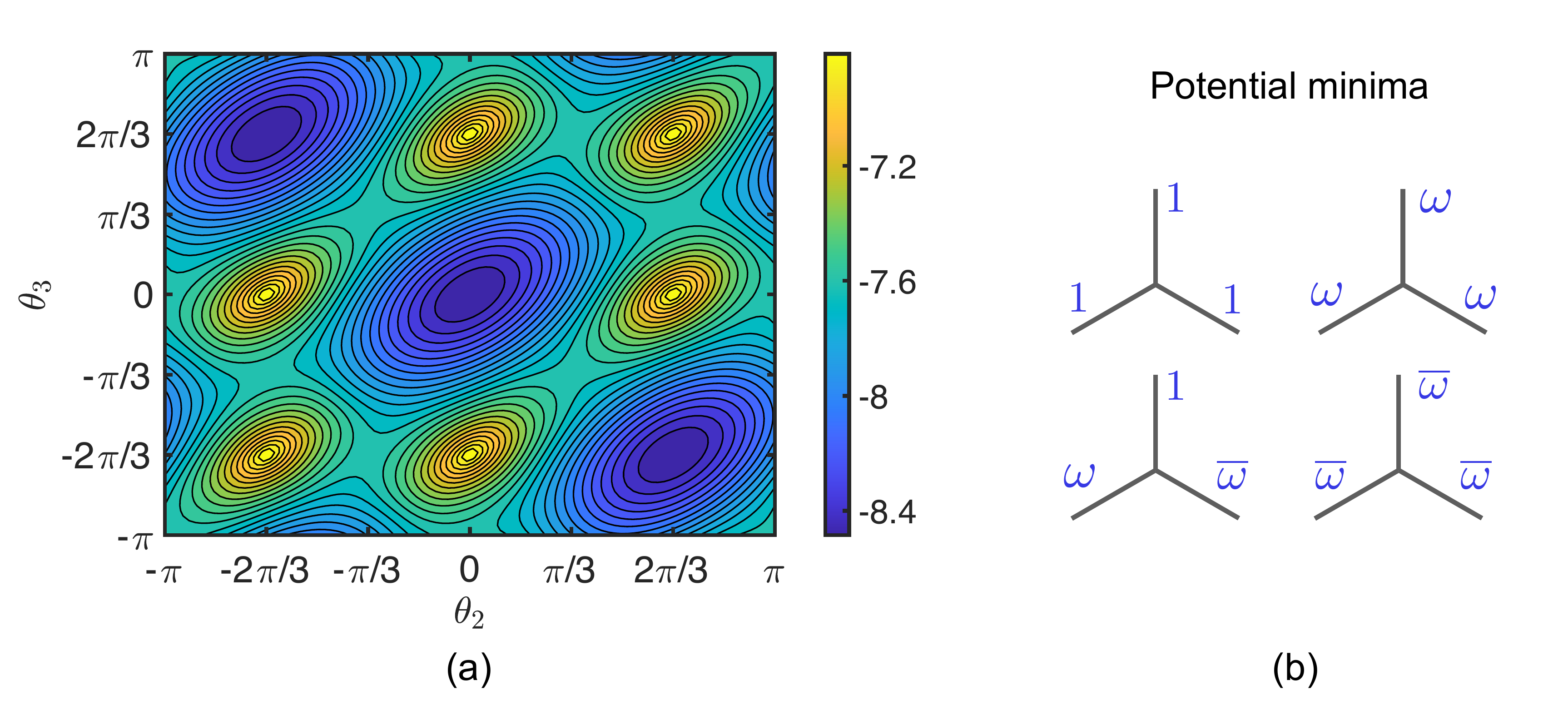}
\caption{(a) Contour plot of the potential $-2 \sum_a \left|\sum_i W_{ai} e^{i\theta_i} \right|$ as a function of $\theta_2$ and $\theta_3$. We fix $\theta_1=0$. (b) All inequivalent gauge wire phase configurations $(\theta_1, \theta_2, \theta_3)$ at the potential minima. The potential minima satisfy $\prod_i e^{i\theta_i}=1$, corresponding to the usual (uninverted) star term.}
\label{fig:potential_uninverted}
\end{figure*}

We give an explicit construction of the usual $\mathbb{Z}_3$ quantum double with a star term favoring $A_s=1$. Consider the Josephson energy~(\ref{eq:H_J}) with the following $W$ matrix:
\begin{equation}
W = \frac{1}{\sqrt{6}}
\begin{pmatrix}
1 & 1 & \omega \\
1 & \omega  & 1  \\
1 & \overline{\omega}  & \overline{\omega}  \\
1 & 1 & \overline{\omega}  \\
1 & \overline{\omega} & 1 \\
1 & \omega & \omega
\end{pmatrix}.
\label{eq:W_uninverted}
\end{equation}
The corresponding superconducting wire array now contains 6 matter wires and 3 gauge wires per lattice site (“waffle”).  The above $W$ matrix satisfies $W^\dagger W = \mathbb{1}$, and has the following automorphism
\begin{equation}
L^\dagger \ W \ R = W,
\end{equation}
where $L$ and $R$ are monomial matrices. Hence, the Josephson energy is invariant under transformations~(\ref{eq:RL}) on $\theta_i$ and $\phi_a$. We again restrict $R$ to be diagonal matrices that do not change the product of the three gauge wire phases $\prod_i e^{i\theta_i}$ at each vertex. For example, if we take a $R$ matrix 
\begin{equation}
R = 
\begin{pmatrix}
1 & 0 & 0 \\
0 & \omega & 0 \\
0 & 0 & \overline{\omega}
\end{pmatrix},
\end{equation}
and a $L$ matrix
\begin{equation}
L = 
\begin{pmatrix}
0 & 1 & 0 & 0 & 0 & 0 \\
0 & 0 & 1 & 0 & 0 & 0 \\
1 & 0 & 0 & 0 & 0 & 0 \\
0 & 0 & 0 & 0 & 0 & 1 \\
0 & 0 & 0 & 1 & 0 & 0 \\
0 & 0 & 0 & 0 & 1 & 0
\end{pmatrix},
\end{equation}
it is easy to check that the automorphism~(\ref{eq:auto}) holds. In Fig.~\ref{fig:potential_uninverted}(a), we plot the potential energy profile Eq.~(\ref{eq:minE}) for the $W$ matrix~(\ref{eq:W_uninverted}), which shows the minima of the Josephson energy. We find that the minima now correspond to gauge wire phases with a zero net flux $\prod_i e^{i\theta_i}=1$. In Fig.~\ref{fig:potential_uninverted}(b), we show all inequivalent gauge wire phase configurations at the potential minima. Therefore, the effective $\mathbb{Z}_3$ description of the star term now has the usual form: $-J \sum_s (A_s + A_s^\dagger)$ with $J>0$.


\bibliography{reference}


\end{document}